\documentclass[fleqn,usenatbib]{mnras}
\usepackage[T1]{fontenc}

\usepackage{newtxtext,newtxmath}

\usepackage{ae,aecompl}
\usepackage{graphicx}	
\usepackage{amsmath}	
\usepackage{bm}	
\usepackage{hyperref}
\newcommand{\hMsun}{ h^{-1}{\rm M_{ \odot}}}
\newcommand{\hMpc}{ h^{-1}{\rm Mpc}}

\newcommand{\ihMpcC}{ h^{3}{\rm Mpc}^{-3}}

\newcommand{\ihMpc}{ h\,{\rm Mpc}^{-1}}

\newcommand{\vpeak}{V_{\rm peak}}
\newcommand{\mpeak}{M_{\rm peak}}
\newcommand{\fkcen}{f_k^{\rm cen}}
\newcommand{\fksat}{f_k^{\rm sat}}
\title[Assembly bias in SHAM]{A flexible modelling of galaxy assembly bias}

\author[S. Contreras et al.]{
S. Contreras,$^{1}$\thanks{E-mail: sergio.contreras@dipc.org}
R. E. Angulo,$^{1,2}$
 \& M. Zennaro$^{1}$.
\\
$^{1}$Donostia International Physics Center (DIPC), Manuel Lardizabal Ibilbidea, 4, 20018 Donostia, Gipuzkoa, Spain.\\
$^{2}$IKERBASQUE, Basque Foundation for Science, 48013, Bilbao, Spain.
}

\date{Accepted XXX. Received YYY; in original form ZZZ}

\pubyear{2020}

\begin{document}
\label{firstpage}
\pagerange{\pageref{firstpage}--\pageref{lastpage}}
\maketitle

\begin{abstract}
We use the {\sc Illustris TNG300} magneto-hydrodynamic simulation, the {\sc SAGE} semi-analytical model, and the subhalo abundance matching technique (SHAM) to examine the diversity in predictions for galaxy assembly bias (i.e. the difference in the large scale clustering of galaxies at a fixed halo mass due to correlations with the assembly history and other properties of host haloes). We consider samples of galaxies selected according to their stellar mass or star formation rate at various redshifts. We find that all models predict an assembly bias signal of different magnitude, redshift evolution, and dependence with selection criteria and number density. To model these non-trivial dependences, we propose an extension to the standard SHAM technique so it can include arbitrary amounts of assembly bias. We do this by preferentially selecting subhaloes with the same internal property but different {\it individual} large-scale bias. We find that with this model, we can successfully reproduce the galaxy assembly bias signal in either {\sc SAGE} or the {\sc TNG}, for all redshifts and galaxy number densities. We anticipate that this model can be used to constrain the level of assembly bias in observations and aid in the creation of more realistic mock galaxy catalogues.
\end{abstract}


\section{Introduction}
\label{sec:Introduction}
The basis of modern galaxy formation theory was laid down by \cite{WhiteRees:1978}, who proposed that galaxies form and evolve inside dark haloes. The field of galaxy formation has progressed enormously since then, but this basic premise still holds. A natural corollary is that the properties of galaxies should be intimately related to the properties of their host haloes. 

This fundamental galaxy-halo connection is at the heart of several of the most popular models currently used to interpret galaxy clustering measurements. One of these popular techniques is the ``halo occupation distribution'' (HOD) which describes the abundance of galaxies inside a given halo as a parametric function of the host halo mass. HOD dates back to the early 2000s (e.g.,  \citealt{Jing:1998a,Benson:2000,Peacock:2000,Seljak:2000,Scoccimarro:2001,Berlind:2002,Berlind:2003,Cooray:2002,Yang:2003}), and even today it is routinely employed to interpret observations of the correlation function of galaxies, infer the typical halo masses of observed galaxies, build mock catalogues, and even constrain cosmological parameters (eg. \citealt{AEMULUS3}). 

Another popular method is the so-called subhalo abundance matching (SHAM, e.g. \citealt{Vale:2006,Conroy:2006}), where, essentially, the most massive/luminous galaxies are assumed to be hosted by the most massive subhaloes. SHAM variants have shown to be remarkably accurate in reproducing the clustering of galaxies in observations \citep{Reddick:2013} and in hydrodynamical simulations \citep{ChavesMontero:2016}. These models have recently evolved into more sophisticated empirical models which attempt to interpret a wide range of galaxy properties \citep{Moster:2018,Behroozi:2019}.

In the context of increasingly accurate galaxy surveys and clustering measurements, one of the main limitations of these models is the amount of ``galaxy assembly bias'' they predict. The galaxy assembly bias is the excess (or lack of) large scale clustering of a galaxy sample caused by details of how the galaxy-halo connection depends on halo assembly history and properties other than mass. This concept must not be confused with the differences in the clustering of galaxies with the same halo mass but different secondary halo property, which is technically halo assembly bias traced by galaxies and not galaxy assembly bias, see \citealt{Croton:2007} for more details. Galaxy assembly bias is the consequence of two effects: halo assembly bias and occupation variation. Halo assembly bias (e.g. \citealt{Gao:2005}) is the difference in halo clustering among haloes of the same mass but a different secondary property (e.g. formation time, concentration, spin, etc). Occupancy variation \citep[see][]{Zehavi:2018,Artale:2018} is the dependence of the galaxy population on halo properties other than mass. Note that none of these effects on its own would cause galaxy assembly bias.

The degree of galaxy assembly bias predicted by realistic galaxy formation models has been studied by various authors. 
\cite{C19} found that the level of galaxy assembly bias increases with number density and decreases with redshift for both, stellar mass and SFR-selected samples. These authors found also that the amplitude is always higher for stellar mass-selected samples (with $\sim 15\%$ and $\sim 3\%$ of galaxy assembly bias signal for the stellar mass and SFR selected sample respectively, at its higher value) and that it can become negative for the most extreme cases (e.g. to be $\sim 10\%$ negative for a sample with $n = 0.001 \ihMpcC$ at z=3). \cite{ChavesMontero:2016} detected galaxy assembly bias to be of 15\% in the EAGLE hydrodynamical simulation at $z=0$ for various stellar-mass selected samples. More recently, \cite{Montero-Dorta:2020} showed that galaxies in the Illustris-TNG simulation cluster differently depending on the properties of their host haloes.

Since in the standard HODs the galaxy population of a halo depends only on its halo mass,  their predictions have no galaxy assembly bias. On the contrary, the abundance of SHAM galaxies does depend on the halo assembly (e.g. recently-formed haloes host more subhaloes), and it predicts that about 10\% of the galaxy clustering to be caused by galaxy assembly bias. Note, however, that in general the amount of assembly bias is expected to be connected with environmental processes (e.g. \citealt{Dalal:2008,Ramakrishnan:2019,Mansfield:2020} which might or might not be captured accurately in current hydrodynamical simulations and galaxy formation models. This suggests that if a given observed galaxy population has a different assembly bias to that in the model, the respective inferences about cosmology or galaxy formation will be biased.

This problem has motivated several attempts to incorporate assembly bias in HODs \citep{Paranjape:2015}. They have had, however, limited success. One of the most common ways to add assembly bias to empirical techniques is the decorated HOD approach \citep{Hearin:2016}. In this approach, the halo occupation is splited in two-parts (per halo mass bin) depending on a secondary halo property that contains halo assembly bias (e.g. concentration). Then the galaxy occupation of these sub-population is varied to imprint assembly bias on the mock sample, keeping the same mean galaxy occupation. The main issue with this method is the selection of the secondary property. The most commonly used property is halo concentration (e.g. \citealt{Wang:2019,Zentner:2019,Vakili:2019}), which, although contains an amount of halo assembly bias (e.g. \citealt{Gao:2005})  it is only responsible of a small part of the total galaxy assembly bias signal, and so, even considering an unrealistic relation between halo occupation and concentration, by itself is not enough to reproduce the full galaxy assembly bias signal of a galaxy sample (e.g. \citealt{Croton:2007, Hadzhiyska:2020,Xu:2020}). Only a few other works have tried other halo properties, like environment (eg. \citealt{McEwen:2016,Xu:2020}). To our knowledge, the only attempt of modelling assembly bias in SHAM is the work of \cite{Lehmann:2017}, also using concentration as a way to add galaxy assembly bias. Lehmann et al. model also allowed a change in the satellite fraction when varying the level of assembly bias, which not only changes the assembly bias of the sample, but also the total bias.

In the first part of this paper, we aim to systematically quantify how similar or different the predictions for the galaxy assembly bias signal is on different state-of-the-art galaxy formation models. For that, we employ a semi-analytical model, a hydrodynamic simulation, and a SHAM mock. Specifically, we will use the Illustris TNG300 simulation \citep{TNGa}; the SAGE semi-analytical model \citep{Croton:2016}, and SHAM mocks using $\vpeak$ as main subhalo property. All these modellings were carried out over the same simulated cosmic volume, which facilitates their comparison. Here we will find that the galaxy assembly bias signal is not universal and that different models predict very different amplitude, redshift evolution, and dependence with the number density of the sample.

Motivated by this finding, in the second part of this paper we will propose a flexible formalism to include galaxy assembly bias in SHAM. In short, this method adds a tuneable degree of correlation between the large-scale bias of individual subhaloes and the scatter in stellar mass for a fixed $\vpeak$. We will demonstrate that this approach is flexible enough to mimic the galaxy assembly bias as measured in SAGE as well as in the TNG300 catalogues, at all redshifts and number densities. This improves over previous works (eg. \citealt{Lehmann:2017}) by forcing a constant abundance of satellite galaxies. This means that the galaxy assembly bias introduced by our method should not contain other sources of bias, making it easier to interpret the results of our work. The model presented here can be easily extend by using any secondary halo property (eg. environment, halo age, concentration) including the ``object-by-object'' bias \citep{Paranjape:2018}, that is the property we use to give assembly bias to our galaxies. We anticipate that being able to create SHAM samples with any amount of assembly bias would ultimately result in a more accurate interpretation of observational data. 

The outline of the paper is the following: in \S~\ref{sec:models} we describe the three different methods to model galaxies we use. In \S~\ref{sec:GAB} we quantify the magnitude and redshift evolution of the galaxy assembly bias signal in these models. We also explore the causes of the galaxy assembly bias in our samples. In \S~\ref{sec:SHAM_AB} we present a new extension to the SHAM algorithm that enables flexible modelling of the galaxy assembly bias. We finalize in \S~\ref{sec:Conclusions} with our conclusions and a summary of our main results.

\section{Simulations and empirical models}
\label{sec:models}

In this section we first describe the three different galaxy formation models we will analyse: a state-of-the-art hydrodynamical simulation, a semi-analytic galaxy formation model, and an empirical model. We then describe the galaxy samples catalogues we will use throughout this paper. 

\subsection{The TNG300 }
\label{sec:TNG300}

The hydrodynamical simulation we will consider is ``The Next Generation'' Illustris Simulations (IllustrisTNG, \citealt{TNGa, TNGb, TNGc, TNGd, TNGe}). The Illustris-TNG is a suite of magneto-hydrodynamic cosmological simulations, successors of the original Illustris simulation \citep{Illustrisa,Illustrisb,Illustrisc,Illustrisd}. The simulations were run using AREPO \citep{AREPO} adopting cosmological parameters consistent with recent analyses \citep{Planck2015}. Specifically, $\Omega_{\rm dm}$ = 0.3089, $\Omega_{\rm b}$ = 0.0486, $\sigma_8$ = 0.8159, $n_s$ = 0.9667 and $h$ = 0.6774. 

These simulations feature a series of improvements upon their predecessor, the Illustris simulation, including: i) an updated kinetic AGN feedback model for the low accretion state \citep{Weinberger:2017}; ii) an improved parameterisation of galactic winds \citep{Pillepich:2018}; and iii) the inclusion of magnetic fields based on ideal magneto-hydrodynamics \citep{Pakmor:2011,Pakmor:2013,Pakmor:2014}.

In this paper, we will use the Illustris-TNG300 (TNG300 thereafter), which is the largest high-resolution hydrodynamic simulation currently available in the world. This simulated volume is a periodic box of 205 $\hMpc$ ($302.5 \sim 300$ Mpc) aside. The number of dark matter particles and gas cells is $2500^3$ each, implying a baryonic mass resolution of $7.44\times10^6\,\hMsun$ and of $3.98\times 10^7\,\hMsun$ for dark matter.  We will analyse the $z=0$, $z=0.5$,  and $z=1$ outputs, publicly available at the TNG project webpage\footnote{\url{https://www.tng-project.org/}}. 

Note that we will consider catalogues of galaxies with stellar masses above $\sim 5\times10^{9}\,\hMsun$, which means galaxies resolved with more than $1000$ resolution elements. We define the stellar mass of a galaxy as the mass of all star particles bound to the respecive main subhalo. Particles bound to subtructure of a subhalo are not included. The SFR is defined as the sum of the individual instantaneous star formation rates of all gas cells in the subhalo.

In addition to the TNG300, we will also employ its dark-matter-only counterpart, TNG300-Dark, as the basis for our SAM and SHAM models. This gravity-only simulation was carried out with the same initial white-noise field and with the same number of dark matter particles ($2500^3$) as the TNG300, which implies a particle mass of $4.73\times10^7\,\hMsun$. In some parts of our analysis, we will also use the TNG300-2-Dark and TNG300-3-Dark, lower-resolution versions of the TNG300-Dark ran with the same initial conditions but with only $1250^3$ and $625^3$ particles of mass $3.78\times 10^8 h^{-1}M_{\odot}$ and $3.03\times 10^9 h^{-1}M_{\odot}$, respectively. Finally, we will use the respective subhalo merger trees to compute additional subhalo properties such as the peak maximum circular velocity (referred to as $\vpeak$).

\begin{figure}
\includegraphics[width=0.45\textwidth]{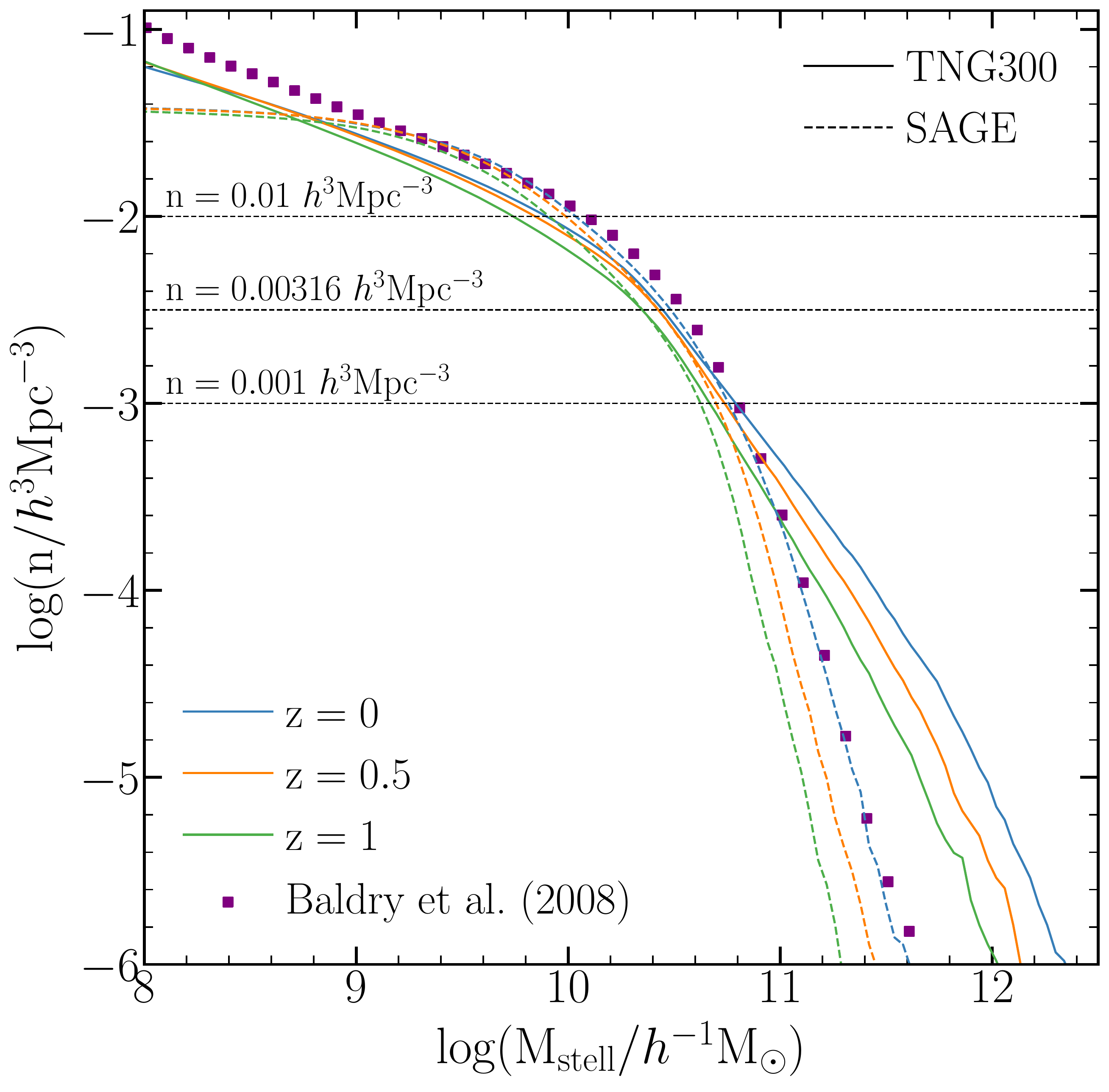}
\includegraphics[width=0.45\textwidth]{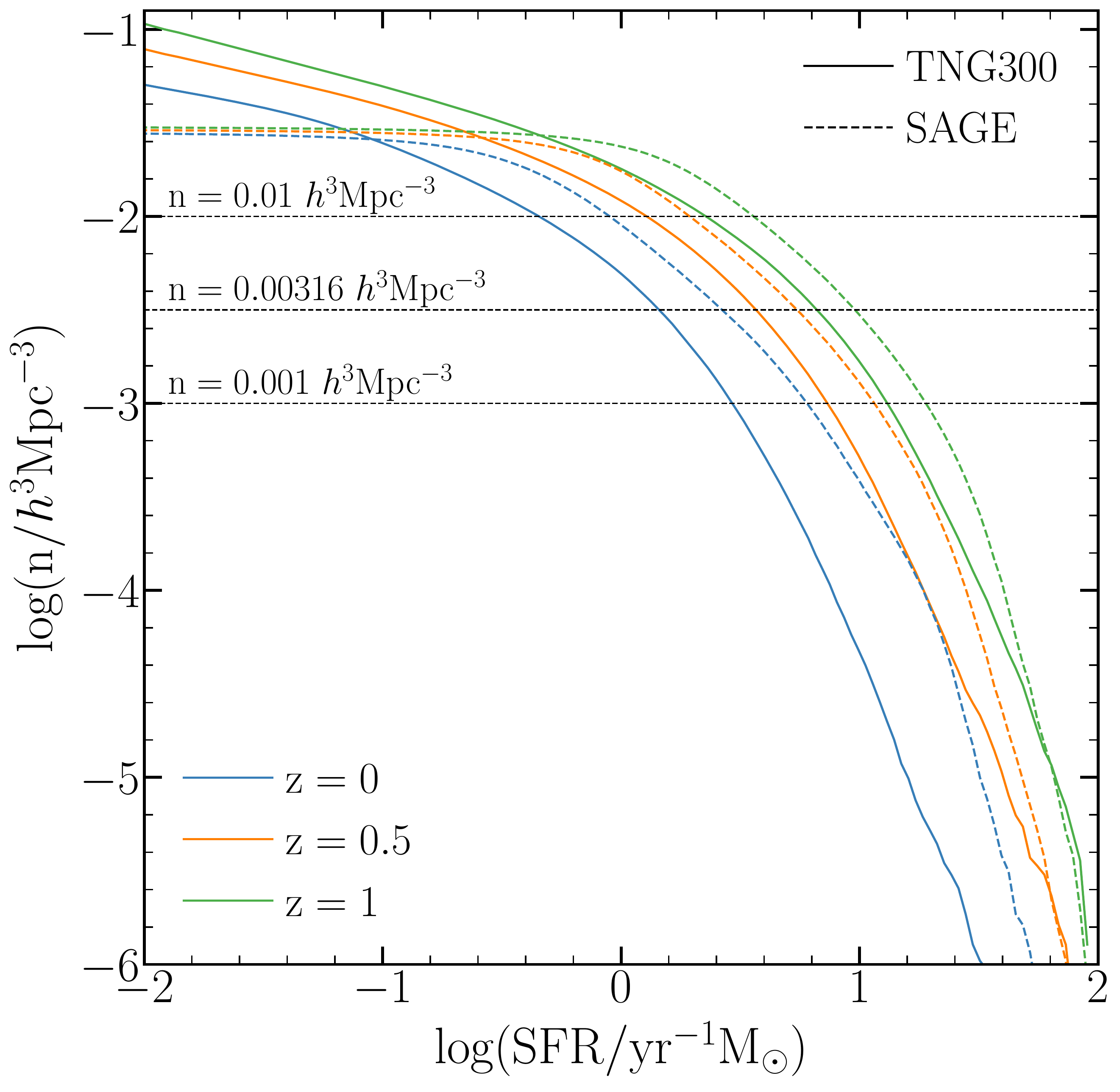}
\caption{The cumulative stellar mass function (top panel) and the cumulative SFR function (bottom panel) predicted by the TNG300 hydrodynamic simulation (solid lines) and the SAGE semi-analytical galaxy formation models (dashed lines). Different colours indicate different redshifts, as labelled. Dotted horizontal lines mark the number densities of the samples used in this work. The galaxies included in a sample are those located to the right of the intersection between the solid or dashed lines and horizontal dotted lines. For comparison, for the cumulative stellar mass function we show the observational data from Baldry et al. (2008).}
\label{Fig:CMF}
\end{figure}

\subsection{SAGE}
\label{sec:SAGE}

The second model we consider is a semi-analytic galaxy formation model (SAM). Specifically, we consider the ``Semi-Analytic Galaxy Evolution'' code \cite[SAGE,][]{Croton:2016}, a SAM based on the model presented in \cite{Croton:2006} and the L-Galaxies code \citep{Henriques:2015}. This model includes a variety of physical processes -- gas cooling, star formation, chemical enrichment, etc -- and was the first galaxy formation model to include feedback from AGNs as a mean of suppressing star formation on massive galaxies (along with \citealt{Bower:2006}). 

One of the main characteristics of this SAM is that it does not use orphan subhaloes, i.e. subhaloes that are not possible to identify in the simulation for numerical reasons, but that are expected to still exist and host a galaxy. This is so the model can be easily run on any dark matter simulation, as long as the merger trees are provided in an appropriate format.

We run SAGE on the merger trees of the TNG300-3-Dark simulation. This simulation has a slightly lower mass resolution than the original simulation employed to calibrate its free parameters (the Millennium Simulation, \citealt{Springel:2005}). We check the main predictions of SAGE finding good agreement with the observed stellar mass function. We also test running SAGE over the TNG300-2-Dark and the TNG300-1-Dark, that have a much higher mass resolution that the Millennium Simulation, finding less agreement with the observed stellar mass function, especially at low masses. 

We expect SAGE to provide numerically robust predictions for the number densities studied here. We use the default calibration of the model to run this SAM. While this could introduce some differences compared to a calibrated SAGE for the specific cosmology of the TNG suite, we found that a new calibration would only introduce differences in the main prediction of the SAM (not shown here).

\subsection{Subhalo abundance matching}
\label{sec:SHAM}

The third model we consider is the so-called ``subhalo abundance matching''. SHAM is an empirical method to populate subhaloes of an $N$-body simulation with galaxies. In its most basic version, SHAM assumes a one-to-one mapping between the mass of a subhalo and its stellar mass or luminosity. More recent implementations of SHAM add scatter to this mapping and include satellite galaxies by using subhalo properties before infall or their maximum values over their full history. These modifications are critical to get even approximately accurate results in agreement with observed clustering. 

One of the main advantages of SHAMs is their predictability while being computationally efficient. In most implementations, they use a single free parameter, the scatter between the subhalo  property used and the stellar mass, in contrast to HOD models which use between 5 and 10 free parameters (if assembly bias, velocity bias and other effects are included). Additionally, SHAM predicts galaxy clustering in rough agreement with hydrodynamical simulations and reproduce some of it galaxy assembly bias signal, but not all \citep{ChavesMontero:2016}.  

In this paper, we use the TNG300-Dark to create our SHAM mocks with $\vpeak$ as the subhalo property. We adopt a scatter of $0.125$ dex between $\vpeak$ and stellar mass, which is set by measuring this value directly in the outputs of the TNG300. We chose $\vpeak$ as the main property of our SHAM because (a) is widely used in the literature, and (b) we find it has a better agreement between centrals as satellite galaxies compared to other properties such as $\mpeak$ \citep[see also the discussion in][]{Campbell:2018}. Finally, we assign a stellar-mass to each subhalo as that of the galaxy in the TNG300 at the same rank in a list sorted by stellar mass.  We also tested using a different stellar mass function for our SHAMs (the one of SAGE-SAM), finding sub per cent differences in the correlation function, when selecting the galaxies according to a fixed number density. This means that our conclusions should be independent of the mass function selected.

\begin{figure}
\includegraphics[width=0.45\textwidth]{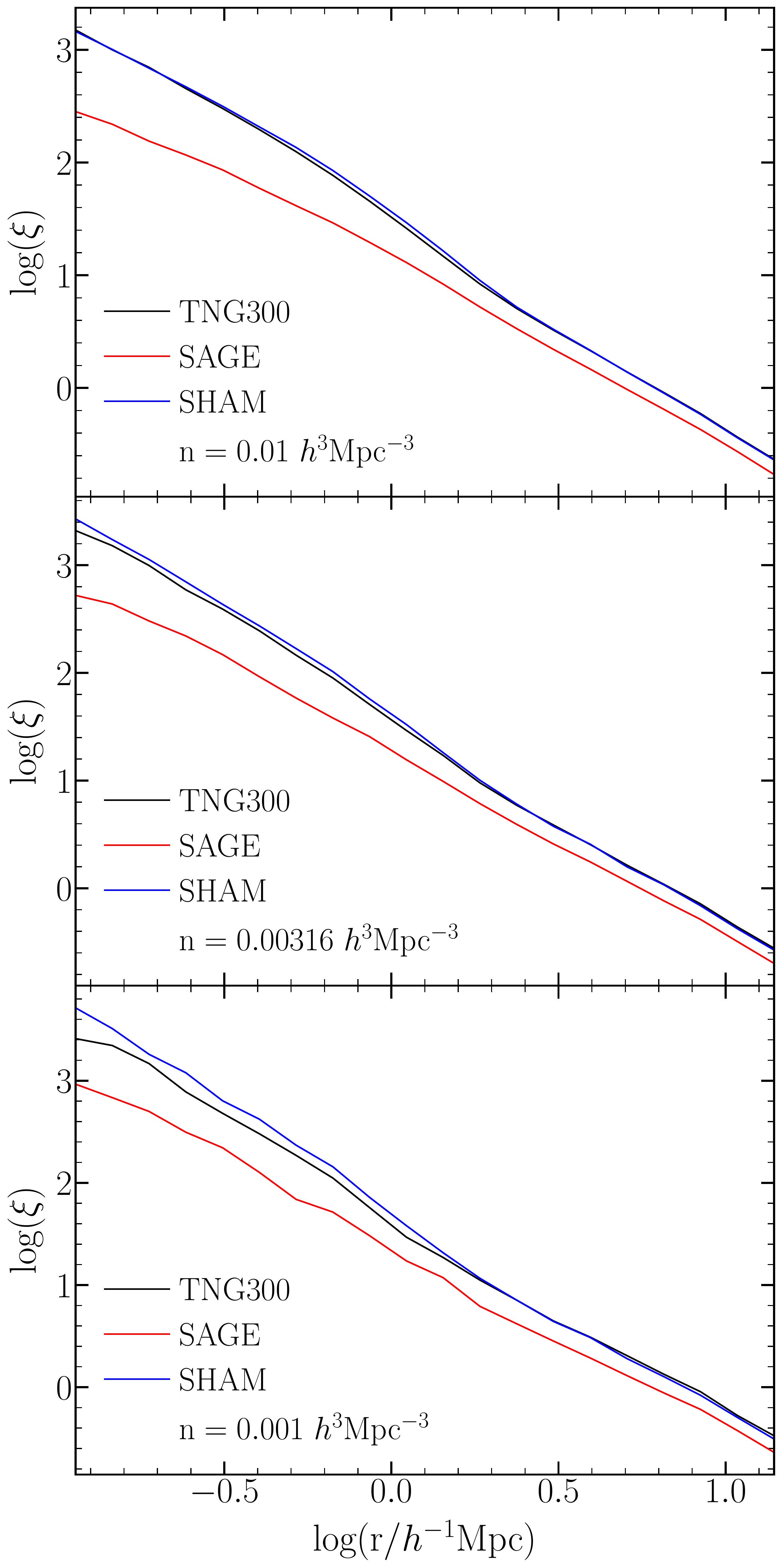}
\caption{The correlation functions of galaxies at $z=0$ in the TNG300 simulation (black solid lines), SAGE semi-analytical model (red solid lines), and SHAMs mocks (blue solid lines). Top, middle, and bottom panels show the predictions for three samples selected by stellar mass with number densities of $\rm n=0.01,\ 0.00316,\ \&\ 0.001\ \ihMpcC$, roughly corresponding to stellar mass cuts of $6\times10^9$, $2\times10^{10}$, and $5\times10^{10}\,\hMsun$, respectively.}
\label{Fig:xi_SHAM}
\end{figure}

\subsection{Galaxy catalogues}

In the following sections, we will measure and compare the assembly bias signal predicted in the three models described before. For each model we will consider three galaxy samples selected according to either star formation rate or stellar mass and with number densities of $n=0.01\,\ihMpcC$, $n=0.00316\,\ihMpcC$, and $n=0.001\,\ihMpcC$. We build catalogues at $z=0$, $z=0.5$, and $z=1$. Given the volume of the TNG300 simulation, these catalogues contain $8.61\times10^{4}$, $2.72\times10^{5}$, and  $8.61\times10^{5}$ objects, from the sparsest to the densest.

In Fig.~\ref{Fig:CMF} we show the cumulative stellar mass function (top) and star-formation rate (SFR) function (bottom) at our three redshifts. For comparison, for the cumulative stellar mass function we show the observational data from \cite{Baldry:2008}. Solid and dashed lines indicate the results from the TNG300 and SAGE models, respectively (note that, by construction, the stellar mass function in SHAM is identical to that in the TNG300). Both models are in a reasonable agreement, except for the abundance of the most massive/star-forming galaxies (which is sensitive to the definition of how exactly stellar mass is computed, as \cite{TNGd} showed). For SAGE, when comparing with observations we find a good agreement at all masses. For the case of the cumulative SFR function, the difference between the models are consistent with those reported by \cite{C13,C15}, who showed that, in general, different galaxy formation models tend to not agree in their predictions for SFRs. Nevertheless, all these discrepancies do help in our aim to explore the variety of predictions from current galaxy formation models. 

Horizontal dotted lines indicate the number density of the three catalogues we will use in this work. By choosing a fixed number density instead of a cut in stellar mass or SFR, we facilitate the comparison with other galaxy formation models/mocks that do not share the same stellar mass and/or SFR distribution. We summarise the cuts on stellar mass for the different redshift and number densities in Table~\ref{Table:n_cuts}.

\begin{table}
  \caption{The cuts of stellar mass and SFR for the TNG300 and SAGE at the different number densities and redshift output used on this work. The units are $h^{-1}{\rm  M_{\odot}}$ for the stellar masses, $\rm M_{\odot}/yr$ for the SFR and $ h^{-3}{\rm Mpc^3}$ for the densities. 
  \label{Table:n_cuts}}
 \begin{tabular}{c c c c} 
 \hline
   & $n=0.001$ & $n=0.00316$ & $n=0.01$ \\ [0.5ex] 
 \hline
 TNG300 $\rm log(M_{stell})$ &  &  &  \\ 
z = 0   & 10.81 & 10.47 & 9.92 \\
z = 0.5 & 10.78 & 10.44 & 9.86 \\
z = 1   & 10.69 & 10.37 & 9.76 \\
 \hline
 TNG300 $\rm SFR$ &  &  &  \\ 
z = 0   & 3.03 & 1.49 & 0.47 \\
z = 0.5 & 7.59  & 3.81 & 1.32 \\
z = 1   & 13.57  & 6.87 & 2.35 \\
 \hline
 SAGE $\rm log(M_{stell})$ &  &  &  \\ 
z = 0   & 10.78 & 10.50 & 10.06 \\
z = 0.5 & 10.72 & 10.45 & 10.01  \\
z = 1   & 10.60 & 10.28 & 9.75 \\
 \hline
 SAGE $\rm SFR$ &  &  &  \\ 
z = 0   & 6.24 & 2.74 & 0.93 \\
z = 0.5 & 10.64 & 10.37 & 9.93 \\
z = 1   & 19.63 & 9.91 & 3.72 \\ 
 \hline
\end{tabular}
\end{table}

The predicted $z=0$ clustering of our catalogues is shown in Fig.~\ref{Fig:xi_SHAM}. Each panel presents the results for a different number density for our galaxy formation model, as indicated by the legend. There is an overall good agreement between the TNG300 and the SHAM models, with small but systematic differences. The SAGE model tends to underpredict the clustering compared to these two models.

Since all models employ an identical simulated volume, we expect the differences to originate from the galaxy modelling and assumptions, rather than from statistical fluctuations. For instance, SHAM might overestimate the clustering on small scales compared to the other models because using $\vpeak$ is equivalent to assuming that the stellar mass of the objects never decreases (i.e. that there is no stellar stripping). Also, the scatter of SHAM was chosen to mimic that of the TNG300, so it is expected to have a clustering similar to this model. On the other hand, the tendency of SAGE to underpredict the clustering, especially at small scales, could be because of different assumptions for satellite disruption.

Also, by looking at the halo occupation distribution of these models (not shown here) we noticed that SAGE tends to populate less massive (i.e. lower bias) haloes compared to the TNG300 and SHAM, resulting in the difference on large scales.

Therefore, the differences in the clustering are likely to be caused by differences in the {\it physical} assumptions -- e.g. star formation prescription, tidal disruption or quenching that affect the galaxies in this model. Hence, by investigating the galaxy assembly bias signal in these catalogues, we will estimate to which degree its magnitude is a generic prediction of galaxy formation or, instead, what the plausible range of values is. We turn to this question in the next section.

\section{The galaxy assembly bias in the galaxy models}
\label{sec:GAB}

The concept of ``assembly bias'' was first introduced by \cite{Sheth:2004} and \cite{Gao:2005}, and it refers to the dependence of the large-scale clustering of haloes on formation time. This effect was generalised by \cite{Gao:2007} to show that the large-scale halo bias also depends on other secondary properties beside the formation time (as concentration, spin, number of substructures) \citep[see also][]{Wechsler:2006,Faltenbacher:2010} and by \cite{Angulo:2009} to higher-order bias parameters. More recent works have extended this list of secondary properties even further (eg. \citealt{Mao:2018}). The existence of ``halo assembly bias'' in simulated haloes is nowadays widely accepted.

Since the evolution of galaxies and haloes are linked, it is expected that an effect analogous to halo assembly bias exists for galaxies. In fact, this effect was detected by \cite{Croton:2007} in SAMs and it is commonly known as ``galaxy assembly bias''. Specifically, Croton et al. showed that, for a fixed cut in stellar mass, the large-scale clustering of galaxies in SAMs was $10\%$ to $20\%$ higher than that of a sample where the galaxy population was only a function of its host halo mass. Likewise, \cite{ChavesMontero:2016} measured a similar amplitude for the ``galaxy assembly bias'' in stellar-mass selected samples of galaxies in the hydrodynamical simulation EAGLE \citep{Schaye:2015}. The same authors, however, reported that SHAM galaxies had a significantly lower amount of assembly bias. Finally,  \cite{C19}, found that for stellar mass and SFR-selected samples, galaxy assembly bias in SAMs decreased at lower number densities and higher redshifts, even becoming negative. 

Observationally, the situation is even less clear with multiple claims of detection (e.g., 
\citealt{Berlind:2006,Yang:2006,Cooper:2010,Wang:2013b,Lacerna:2014a,Lacerna:2014b,Hearin:2015,Miyatake:2016,Saito:2016,Obuljen:2020}) and non-detection/detection due to different systematics (e.g. \citealt{Campbell:2015b,Zu:2016b,Zu:2017,Busch:2017,Sin:2017,Tinker:2017a,Lacerna:2017}) of assembly bias. In other cases, more data are required to reveal the nature of the reported signal (e.g., \citealt{Montero-Dorta:2017b,Niemiec:2018}). The lack of a clear theoretical expectation has certainly been a difficulty, as it does not provide a clear target nor an optimal observational strategy. An efficient observational strategy to measure assembly bias is key since the predicted halo masses in observation, commonly used to infer the assembly bias signal, is normally biased and highly inaccurate.

In this following section, we will quantify the amplitude of assembly bias as a function of redshift, selection criteria, and number density for catalogues constructed in our three galaxy models; the TNG300 hydrodynamical simulation, the SAGE semi-analytical model, and SHAM mocks. 
 
\subsection{The galaxy assembly bias evolution}

\label{sec:GAB_ev}
\begin{figure*}
\includegraphics[width=0.8\textwidth]{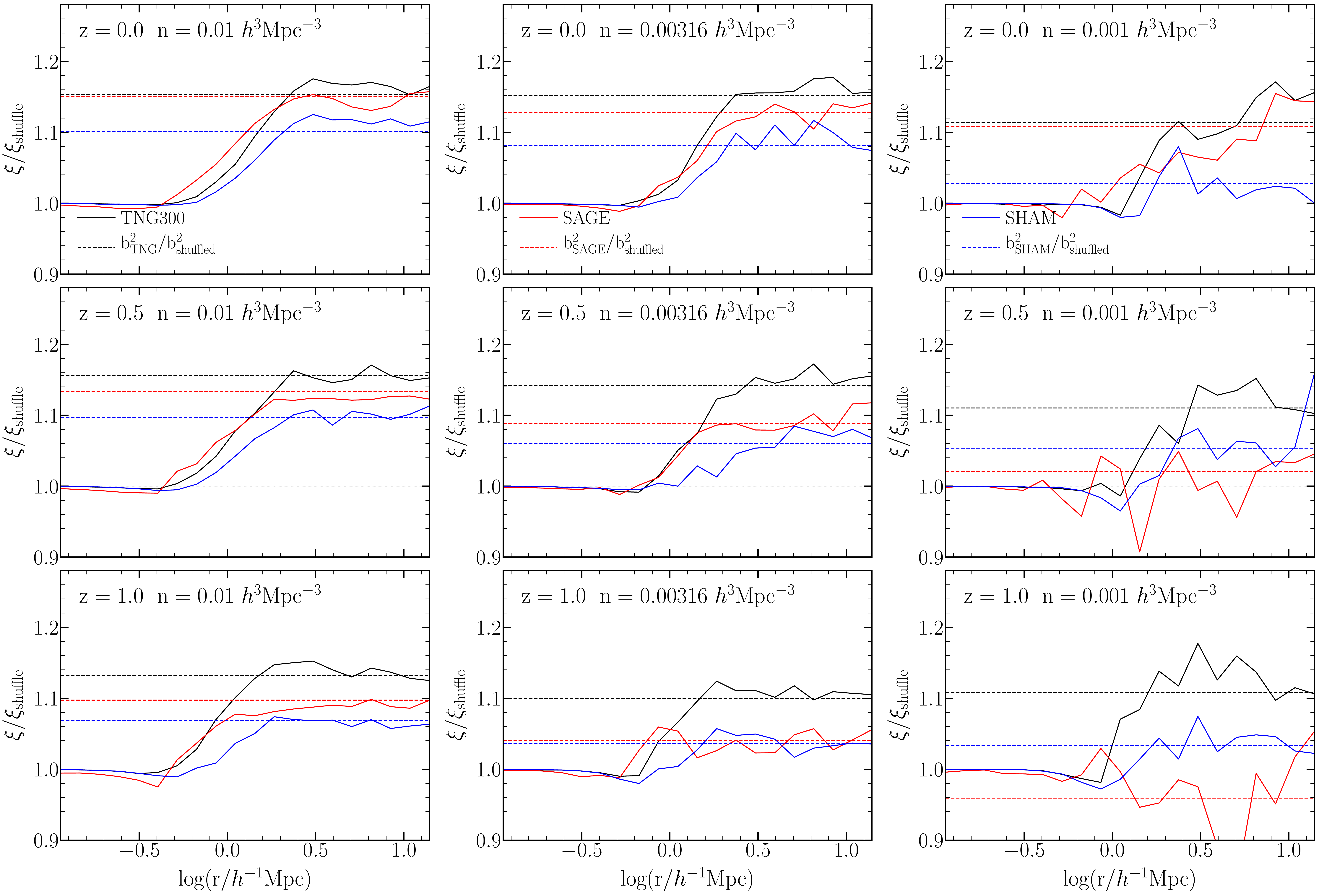}
\caption{The ratio between the correlation functions of the TNG300 simulation (black solid lines), SAGE semi-analytical model (blue solid lines) and SHAMs mocks (red solid lines) with their respective shuffle realisations (i.e. the square of the amount of the galaxy assembly bias signal). The shuffle correlation functions are measured averaging 20 different realisations. The top, middle and bottom rows show the prediction for $\rm z=0,\ z=0.5,\ \&\ z=1$. The left, middle and right column show the predictions for a number density of $\rm n=0.01,\ 0.00316,\ \&\ 0.001$ $h^{3}{\rm Mpc^{-3}}$ for stellar mass-selected galaxies. The dashed horizontal lines show the galaxy assembly bias signal predicted by measuring the individual bias of all the galaxies of each sample, as explained in Section~\ref{sec:GAB_or}.}
\label{Fig:shuffle}
\end{figure*}

\begin{figure*}
\includegraphics[width=0.8\textwidth]{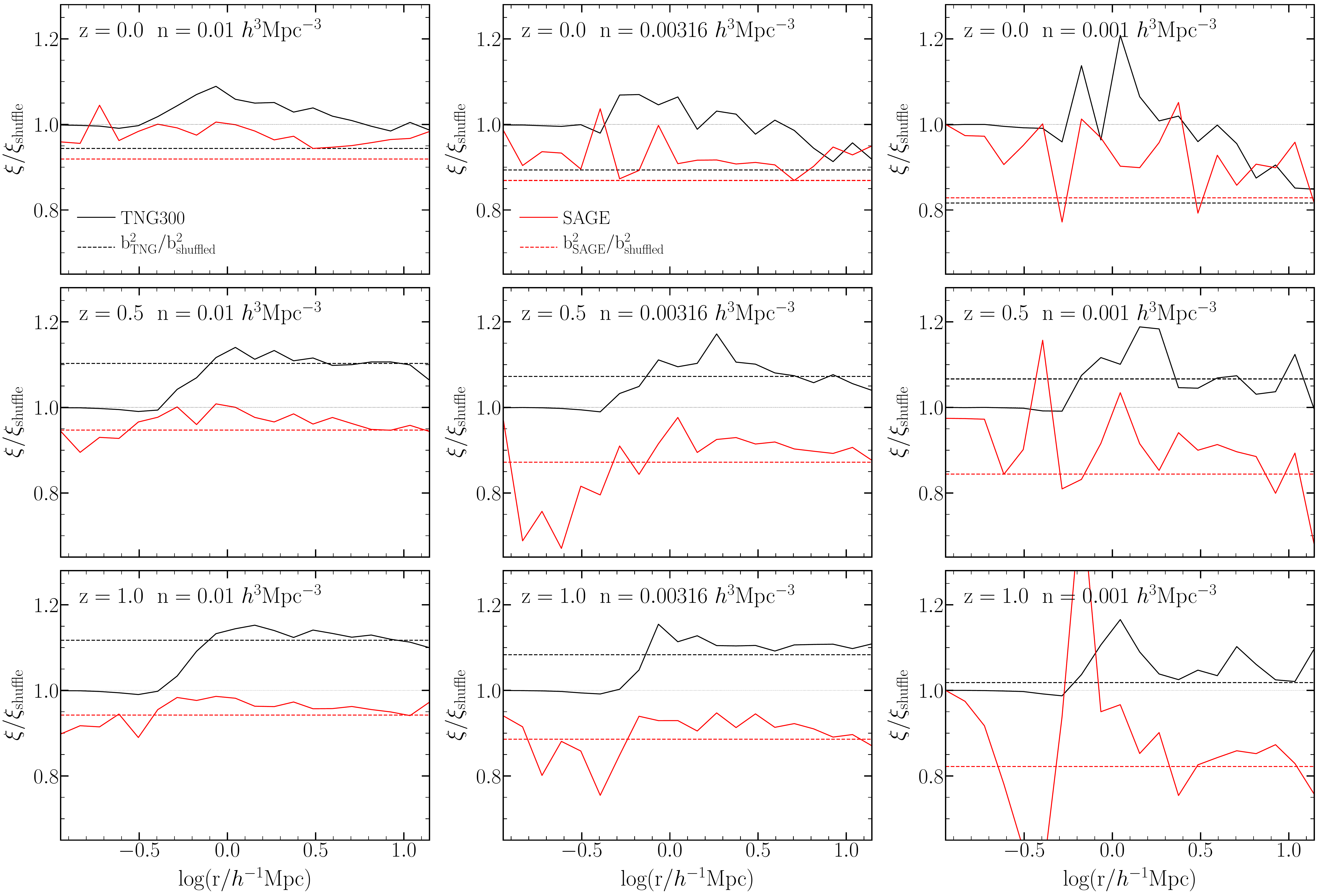}
\caption{Similar to Fig.~\ref{Fig:shuffle}, but for SFR-selected galaxies. We show only the predictions of the TNG300 simulation (black solid line) and SAGE (red solid line), since the standard SHAM implementation does not predict SFR. Note the SAGE curves do not all appear to go to unity on small scales owing to their typical 1-halo term being located on very small scales.}
\label{Fig:shuffle_SFR}
\end{figure*}

\label{sec:GAB_or}
\begin{figure*}
\includegraphics[width=0.8\textwidth]{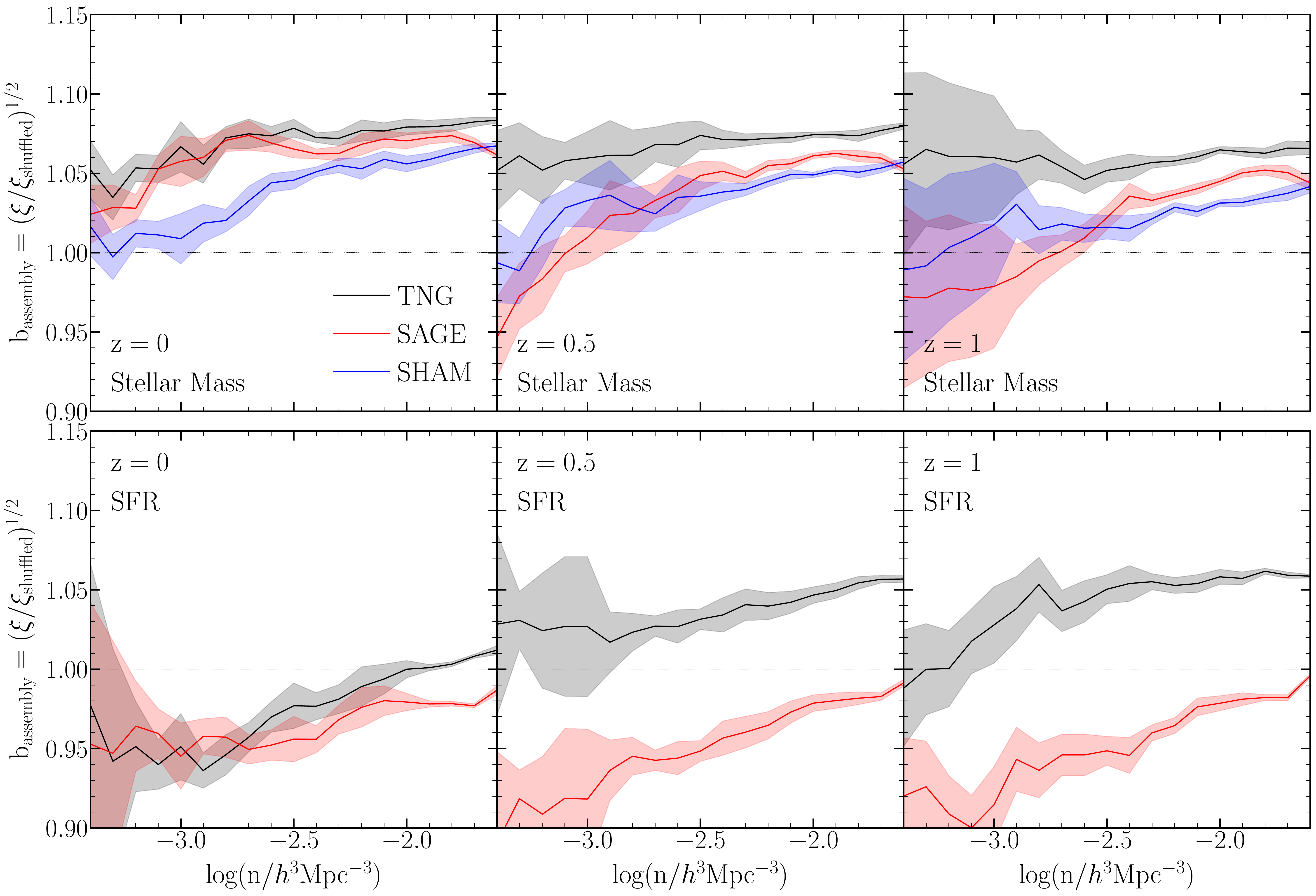}
\caption{The galaxy assembly bias signal of stellar mass-selected galaxies (top panel) and SFR selected galaxies (bottom panel) for z=0 (left panel), z=0.5 (middle panel) and z=1 (right panel). The signal is computed as the square root of the ratio between the correlation function to that of its shuffled counterpart, as shown in Figs.~\ref{Fig:shuffle} and \ref{Fig:shuffle_SFR}, for scales  $3 < r[\hMpc] < 16$. The shaded region represents the standard deviation of the ratio at different scales.}
\label{Fig:bias_ev}
\end{figure*}

{To measure the galaxy assembly bias, $b_{assembly}$}, we compute the ratio between the two-point correlation function of galaxies, $\xi(r)$, to that after randomly shuffling the galaxy population of haloes in mass bins of 0.1 dex following the procedure presented in \cite{Croton:2007}. To reduce stochastic noise, we will display the results after averaging 20 different shuffled catalogues.

There are a few technical details worth highlighting regarding the shuffling procedure. First, we consider haloes in the shuffling even if they do not contain any galaxy. Second, satellite galaxies keep their relative distance to the central galaxy, and the central galaxy is located on the original position of the central galaxy that used to populate that halo. Finally, if there is no galaxy, then we use the position of the centre of the potential of the target halo. 

In Fig.~\ref{Fig:shuffle} we show the ratio between the correlation function of stellar-mass selected catalogues to that of their respective shuffled version. The top, middle and bottom rows show the predictions for $z=0$, $0.5$, and $1$, respectively. Left, middle and right column show the results for number densities, as indicated by the legend. We recall that the magnitude of assembly bias is equal to the square root of the differences on large scales shown here. 

Overall, we can see that all models predict a different amount of assembly bias, different redshift dependence, and different dependence with number density. For instance, TNG300 shows a roughly constant assembly bias signal of $\sim15\%$ in this range of galaxy number density and redshift. On the other hand, SAGE roughly agrees with the TNG300 for $z=0$ at all number densities, but it predicts significantly less at higher redshifts. SHAM, instead,  predicts significantly less assembly bias than SAGE or TNG300, at most redshifts and number densities.  

The differences between models are even larger for SFR-selected galaxies, which is shown in Fig.~\ref{Fig:shuffle_SFR}. Note that we only display results for SAGE and TNG since, in its basic form, SHAM does not predict star formation rates. In this figure we can appreciate that, unlike for the stellar mass selection, SAGE and the TNG300 do not agree on the magnitude of the assembly bias for almost any case. Specifically, the assembly bias signal is much higher for the TNG300 than for SAGE, and it displays a different redshift evolution: the signal slightly decreases with redshift for SAGE, and it significantly increases for the TNG300. We note that the evolution of the signal for SAGE is in similar to that found by \cite{C19} using the \cite{Guo:2013a} SAM.

We summarise these results in Fig~\ref{Fig:bias_ev} which shows the evolution of the assembly bias, as a function of the number density for $z=0$, $z=0.5$ and $z=1$. We compute the assembly bias as the square root of the ratio of the correlation function and its shuffled form, averaged over separations $3 < r/[\hMpc] < 16$. The shaded region indicates one standard deviation of the ratio in each case. 

We would like to emphasise that the amplitude and redshift evolution of the galaxy assembly bias in a given model is a result of the physical processes implemented. For instance, if in a given model galaxies are quenched very rapidly after infall, then a SFR-selected galaxy sample will preferentially select young haloes and would inherit a halo assembly bias. If instead, quenching is very slow, SFR selection would simply return a larger variety of formation times, washing out dependencies with halo formation time. 

These physical processes, and galaxy formation in general, are still very uncertain and many degrees of freedom exist in the (sub-grid) physics implemented, parametric form, as well as in the calibration of the models. This implies that, for the foreseeable future, galaxy assembly bias will not be a prediction of galaxy formation models, but it should rather be considered as an additional parameter to be constrained by models attempting to do inferences from the observed distribution of galaxies. 

Before turning into the problem of incorporating a model for assembly bias in SHAM, in the next section we will investigate and quantify further this effect. 

\subsection{The object-by-object bias}
\label{sec:objectbias}

\begin{figure}
\includegraphics[width=0.45\textwidth]{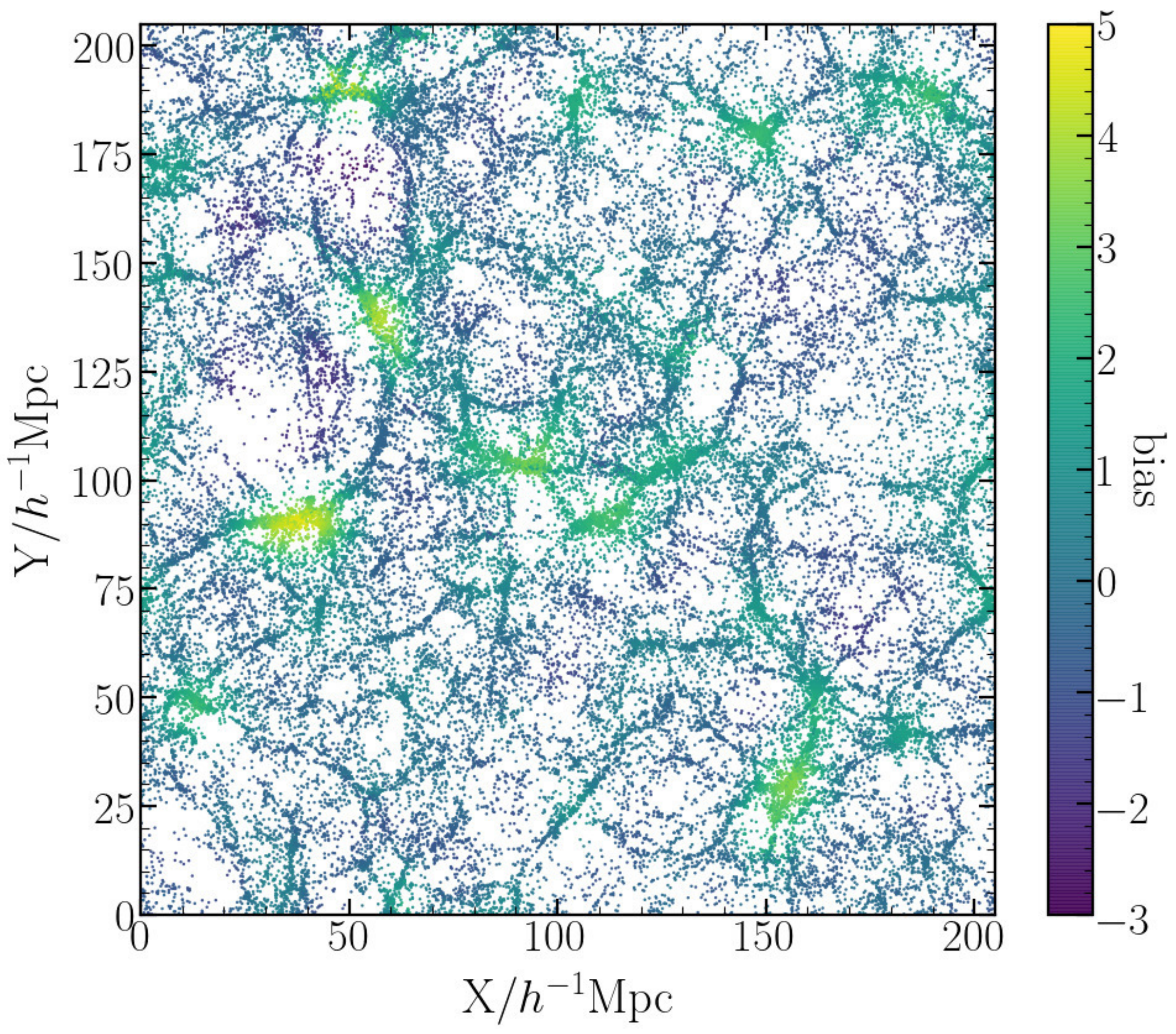}
\caption{The galaxies in a $205\,\hMpc\times\ 205\,\hMpc\times\ 10\,\hMpc$ slice of the TNG300 simulation. The galaxies are colour coded by their individual large-scale bias, as described in \S~\ref{sec:objectbias}. For clarity, only a $10\%$ of the objects are displayed. 
}
\label{Fig:CW}
\end{figure}

To further investigate the galaxy assembly bias in our catalogues, we have computed the large-scale bias of {\it each} galaxy in our sample. We estimate this quantity ``object-by-object'', following \cite{Paranjape:2018} \citep[see also][]{Paranjape:2020}, as:

\begin{equation}
\label{eq:bias}
b_g^i = \left \langle \frac{V}{P(|{\bf k}|) N(k)}  \, \exp(i {\bf k}\cdot {\bf x}^i) \delta^*({\bf k}) \right\rangle_{k},
\end{equation}

\noindent where $V$ is the volume of our simulated box, ${\bf x}$ is the location of a given galaxy, $\delta^*$ is the complex conjugate of the dark matter density field in Fourier space and $P(k)$ its power spectrum. Operationally, we measure $\delta(k)$ from a diluted catalogue of dark matter particles in the TNG300-3 using an NGP assignment scheme on a $256^3$ grid. We have tested that using the TNG300-2 simulation (with $1/2^3$ fewer particles than the original TNG300) gives almost identical results. We perform the average over modes in the range $0.008 < k/{\rm Mpc^{-1}} h < 0.316$. Note that, ideally, we would like to use only scales in the limit $k\rightarrow0$ (e.g. $k/{\rm Mpc^{-1}} h < 0.1$) but given the limited volume of our simulations, we are in the need of using these intermediate scales. Still, we checked that computing the bias using $k/{\rm Mpc^{-1}} h < 0.1$ yields consistent, but noisier, results.  We also tested only computing the ``object-by-object'' bias for the haloes and then assigning it to its substructures, finding identical results.

Intuitively, this estimator corresponds to the cross-correlation between a given point in space and the dark matter density field. Alternatively, it can be regarded as the large-scale overdensity field after a top-hat filter in Fourier space. We highlight that the average of the individual bias of galaxies in a sample is mathematically equivalent to the large scale bias of that sample. 

Fig.~\ref{Fig:CW} shows the distribution of galaxies in a $10\hMpc$ deep slice of the TNG300 catalogue, colour-coded by their individual bias. As expected, galaxies located in denser regions have higher biases than those in less dense regions. Note that galaxies near dense regions, even if they are hosted by low-mass haloes, will still be highly biased.

\begin{figure}
\includegraphics[width=0.45\textwidth]{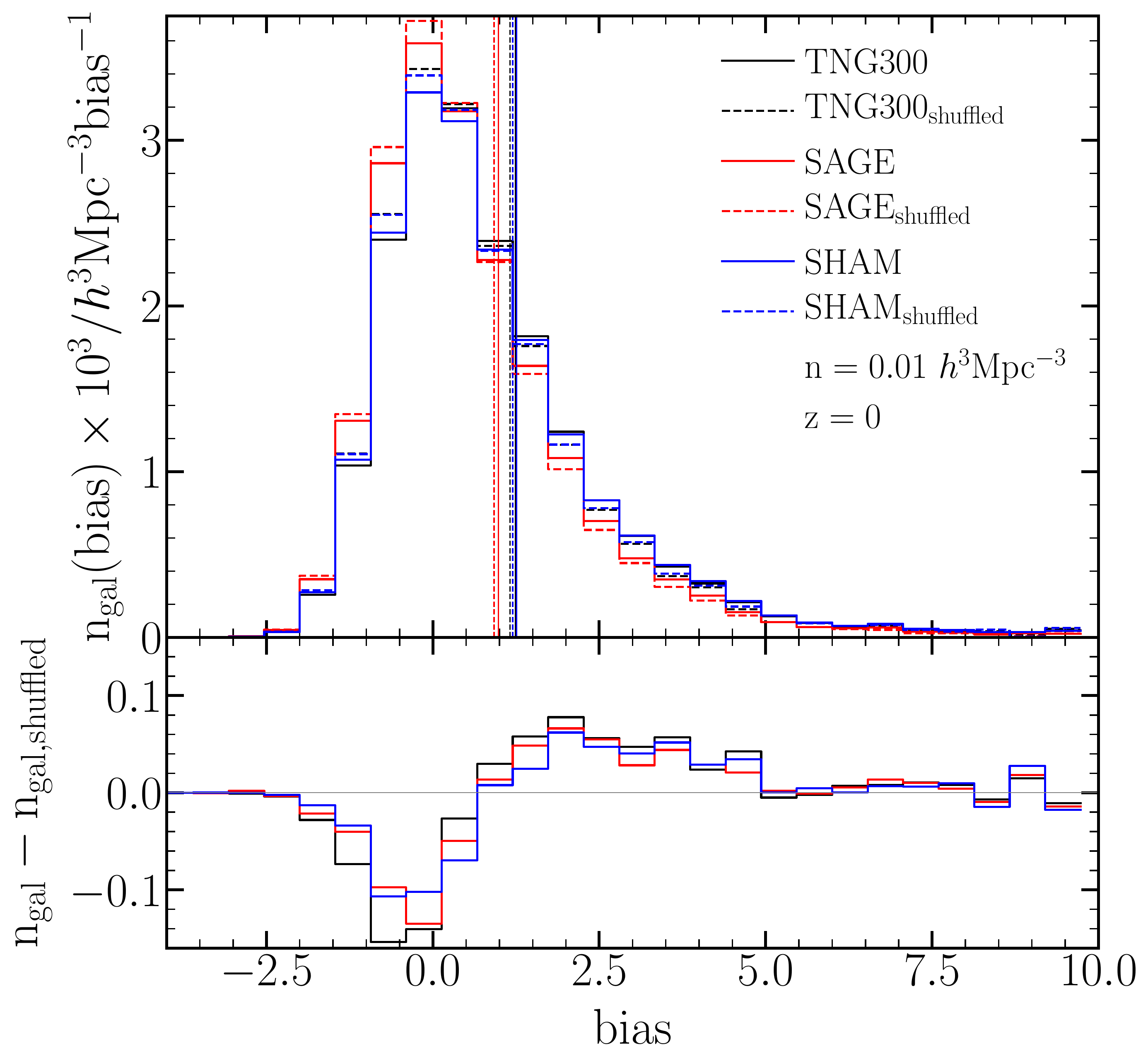}
\caption{The histogram of the individual biases for the galaxies a number density of $\rm n=0.01\ h^3 {\rm Mpc^3}$ selected by stellar mass at z=0. Different colours denote results from our three different galaxy models. In each case, we show the original catalogue and its shuffled version where the galaxy content of a halo is only a function of its mass. Vertical lines mark the average bias in each of the samples. 
\label{Fig:bias_distro}}
\end{figure}

In the top panel of Fig.~\ref{Fig:bias_distro} we show the distribution of individual large-scale biases, $b_g$ (c.f. Eq.~\ref{eq:bias}), for stellar-mass selected galaxies in our three models at $z=0$ and for a number density of $n=0.01\,\ihMpcC$. Solid and dashed lines show the predictions of the original and shuffled samples, respectively. 

Usually, haloes of the same mass are thought to have all the same large-scale bias, which is in fact not true \citep[e.g.][]{Paranjape:2018}. The same holds  for galaxies, as we can see in Fig.~\ref{Fig:bias_distro}, as they can have very different values. The distribution of large-scale biases is very broad: some galaxies have a bias of $\sim10$ whereas others have $-2.5$. This diversity is mostly a consequence of galaxies living in very different environments  -- they can be located in extremely dense regions or in empty voids -- even if hosted by haloes of the same mass. The shuffled version of the catalogue displays a remarkable similar bias distribution. This is because haloes of the same mass also can be found in a wide variety of environments, and also a consequence of halo mass being the primary factor determining the bias of galaxy sample.

 Under a closer inspection, we see that there are systematic differences between the original and shuffled catalogues. This can be better appreciated in the bottom panel, which shows the difference between these two histograms. There we can see that the shuffled sample contains more low-bias galaxies and less high-bias galaxies than the original catalogue. In other words, at a fixed halo mass, the TNG300 simulation preferentially locates galaxies in haloes living in high large-scale density. 

In the case of SHAM, shown as blue curves in Fig.~\ref{Fig:bias_distro}, we see a very similar distribution of individual biases. In particular, the mean (indicated by vertical lines) is almost identical to that in the TNG sample, which is consistent with their clustering agreeing very well (c.f. Fig.~\ref{Fig:xi_SHAM}). In addition, we can see how SHAM also preferentially places galaxies in haloes with higher large-scale bias compared to full halo population. However, this preferential selection is not as strong as in the case of the TNG catalogues. On the other hand, for the case of SAGE we see that even though the average bias is different to that of TNG300 and SHAM (c.f. Fig.~\ref{Fig:xi_SHAM}), the way in which it preferentially selects low and high-bias haloes is more similar to that of the TNG than in the SHAM mocks.

The small difference in the bias distribution implies that the average bias would be slightly different between the original and shuffled catalogues. This is then equivalent to the assembly bias signal! For instance, SAGE and TNG display similar preferences for high/low biased haloes compared to their respective shuffled version. Thus, we expect assembly bias to also be similar. This is in fact what we obtained in Fig.~\ref{Fig:shuffle}. In contrast, we expect SHAM to display less amount of assembly bias, which is also what we found in section \ref{sec:GAB_ev}. 

To see this quantitatively, we have computed the difference of the mean large-scale bias between original and shuffled catalogues for all our samples. Horizontal dashed lines in Figs.~\ref{Fig:shuffle} and \ref{Fig:shuffle_SFR} mark these average values. As we can see, these figures, in fact, coincide remarkable well with the values estimated from the correlation functions. While the match is not perfect in all cases (mostly because of the noisy correlation function measurements) we confirm that the difference in the bias distribution is indeed equivalent to the galaxy assembly bias signal. 

We can, therefore, think of galaxy assembly bias as the consequence of a given model slightly preferring or avoiding regions with different large-scale biases. Different models would have different amounts of ``preferential selection'' that can vary as a function of redshift, selection criteria, etc. Of course, none of the aforementioned models makes an explicit connection between galaxy properties and the large-scale bias. Instead, the underlying physical cause of this can be a mixture of many processes and assumptions in a given galaxy formation model, which correlate with specific details of the halo assembly history, which in turn is correlated with the large-scale overdensity. 

In any case, although the connection between large-scale density and galaxy properties is, in some sense, artificial, this is by definition the galaxy assembly bias. Correlations between galaxy properties and local halo properties, more fundamental from a physics perspective, can at most only partially capture the effect of assembly bias, and are likely to depend sensitively on the underlying galaxy formation physics. A general working model would have needed to consider possible correlations between the occupation number and {\it all} halo/subhalo properties. 

All this suggests an interesting opportunity of using the individual large-scale bias as a second parameter in empirical models. This would open a series of opportunities to search for the origin of the galaxy assembly bias, model observations more precisely, as well as to create mocks with a tunable degree of assembly bias. This should be much more flexible and accurate than other methods that use other secondary properties of the haloes, as the concentration in the decorated HODs \citep{Hearin:2016}, and it would truly cover the full range of assembly bias possible. 

In the remainder of this paper, we will focus on this idea and propose a new version of the subhalo abundance matching that features a tuneable degree of galaxy assembly bias.

\section{Modelling assembly bias in SHAM}
\label{sec:SHAM_AB}

\begin{figure*}
\includegraphics[width=1.05\textwidth]{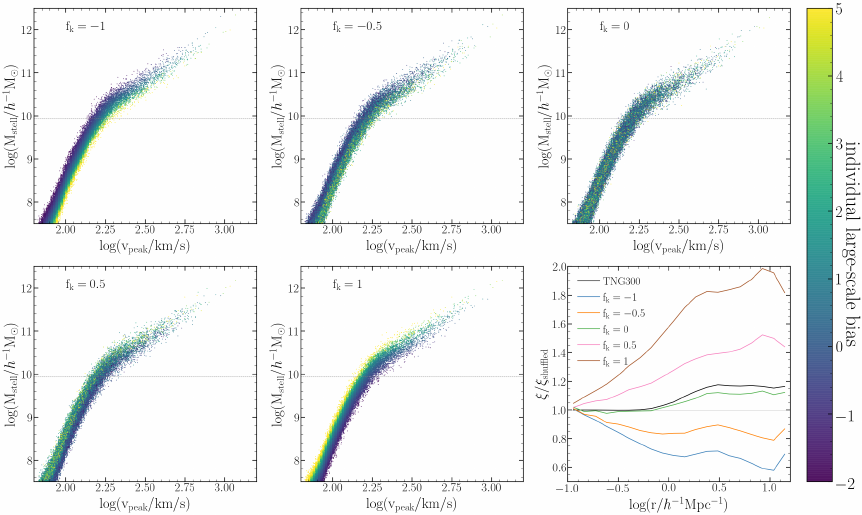}
\caption{The predicted stellar mass of the SHAM as a function of $\vpeak$ for the galaxies of the TNG300 only dark matter at $\rm z=0$. The galaxies are colour coded by their bias (with $f_k$ = $\fkcen$ = $\fksat$), assigned using the implementation descrived in section~\ref{sec:objectbias}. From left to right and top to bottom the galaxies show a perfected anticorrelation between stellar mass and bias ($f_k = -1$), moderated anticorrelation ($f_k = -0.5$), no correlation ($f_k=0$, equivalent to a standard SHAM), moderated correlation ($f_k = 0.5$) and complete correlation ($f_k = 1$). The bottom right panel shows the ratio between the correlation function of these samples and the shuffling run of the standard SHAM.}
\label{Fig:vpeak_mstell_bias} 
\end{figure*}

In the previous sections, we showed that there is not a unique prediction for assembly bias among galaxy formation models, and that this appears very clearly in the correlation between stellar mass and the large-scale bias of each galaxy.

Motivated by this, in this section we propose and test a method to incorporate a tuneable amount of bias in empirical models, such it can mimic the galaxy assembly bias signal from any model or observational sample. Specifically, our method will employ the bias of individual objects as an ancillary parameter to enhance or suppress correlations with the large-scale density field thus providing assembly bias as a degree of freedom. 

Specifically, our idea works in SHAM by re-assigning the stellar mass of galaxies in narrow bins of $\vpeak$ depending on their large-scale bias. This is done for centrals and satellites independently. The stellar mass values of the samples are conserved, only shuffling its values in each bin, meaning the amount of galaxies above a  threshold (eg. the stellar mass cut of our sample) is preserved, as well as the scatter of the sample. Since this is done separately, for centrals and satellites, the satellite fraction of the sample is also preserved. This is done with two correlation parameters, $\fkcen$ and $\fksat$ -- one for central and other for satellite galaxies -- that control the strength of the correlation between the scatter in the $M_*-\vpeak$ and the large-scale bias of each subhalo.    

Different from other methods that use secondary properties on the SHAM to have colours or SFR (e.g. \citealt{Hearin:2013b}) our model changes the intrinsic stellar mass-$\vpeak$ relation, creating an assembly bias signal without the need of employing a secondary galaxy property. In theory, our mocks could be used along with these methods to have colours or SFR more realistically.

Technically speaking, we are not necessarily adding assembly bias to the sample, but just bias. This is because we are using an environmental property to select the haloes, and not a secondary halo property, such as halo concentration, age or spin. Since there is a correlation between several secondary halo properties and environment, we expect that part of this bias can be classified as halo assembly bias, but this is not a requirement for the model. Nevertheless, we will show that this bias has the same behaviour than the galaxy assembly bias from galaxy formation models, and can therefore be used to mimic it.

Now we describe our algorithm. Let us first consider a SHAM sample built using  $\vpeak$ as the primary property. Then, for a given bin in $\vpeak$ and values of $f_k = \{\fkcen, \fksat\}$, our method is as following:

\begin{itemize}
  \item If $f_k > 0$, we sort galaxies in increasing order according to their large-scale bias. Otherwise, we sort the sample in decreasing order. If $f_k = 0$, we randomly reassign the value of the stellar mass.
  \item  We assign a value to each galaxy equal to the ranking in the sorted sample, divided by the number of galaxies in the sample ($g_k$). For example, if $f_k > 0$, and for $N$ galaxies, then the less biased galaxy will have $g_k = 0$, and the most biased will have $g_k = (N-1)/N$.
  \item We define, $f_k^{'} = 1-|f_k|$, and $g_k^{'}$ as a random value between max($g_k-f_k^{'}$, 0) and min($g_k+f_k^{'}$, 1). For example, for $f_k = 0.9$ ($f_k^{'} = 0.1$) a galaxy with a bias equal to the median of the bias of the sample (i.e. $g_k = 0.5$) can have a $g_k^{'}$ between 0.4 and 0.6
  \item Reassign the stellar mass of the galaxies in function of $g_k^{'}$, keeping the same values as the original sample. This means that the galaxy with the largest (lowest) $g_k^{'}$ will have the highest (lowest) stellar mass of the sample. The values available of stellar mass do not change, keeping the same original distribution of stellar masses in the bin of $\vpeak$

\end{itemize}

We repeat this procedure separately for satellite and central galaxies, and for all $\vpeak$ bins. As a result, if $f_k = 0$, then there will be no additional dependence between bias and the stellar mass other than the originally predicted by SHAM. Instead, if $f_k = 1\ (-1)$, then there will be a perfect (anti-) correlation between the large-scale bias and stellar mass for a constant $\vpeak$. If $f_k$ is in between these values, the sample will display different degrees of correlation with the large-scale bias (at a fixed $\vpeak$), and thus it will display different degrees of assembly bias. 

An example of the performance of this method is shown in Fig.~\ref{Fig:vpeak_mstell_bias}. This figure shows the relation between stellar mass and $\vpeak$ present in our SHAM catalogues at $z=0$. We colour-code each galaxy by the large-scale bias. The horizontal dotted line shows the stellar mass cut corresponding to our densest sample. 

In each panel, we show the results after adopting different values for $f_k$, as indicated by the legend. For this particular example, we assume $f_k = \fksat = \fkcen$, i.e. the correlation between the scatter and the large-scale bias to be identical between central and satellite galaxies. We can appreciate that, at a fixed $\vpeak$, negative values of $f_k$ result into a secondary anti-correlation between stellar mass and large-scale bias. On the contrary, positive $f_k$ values imply that at a fixed $\vpeak$ high mass galaxies will be preferentially located in high-bias regions. 

Although not done here, we note that a very similar algorithm could be developed to implement different degrees of assembly bias in HOD models, and in predictions for SFR in SHAM, where the scatter in the predicted star formation rate (e.g. based on the mass accretion rate) could be correlated with the large-scale bias in similar ways as we do for the scatter in stellar mass.

\begin{figure}
\includegraphics[width=0.45\textwidth]{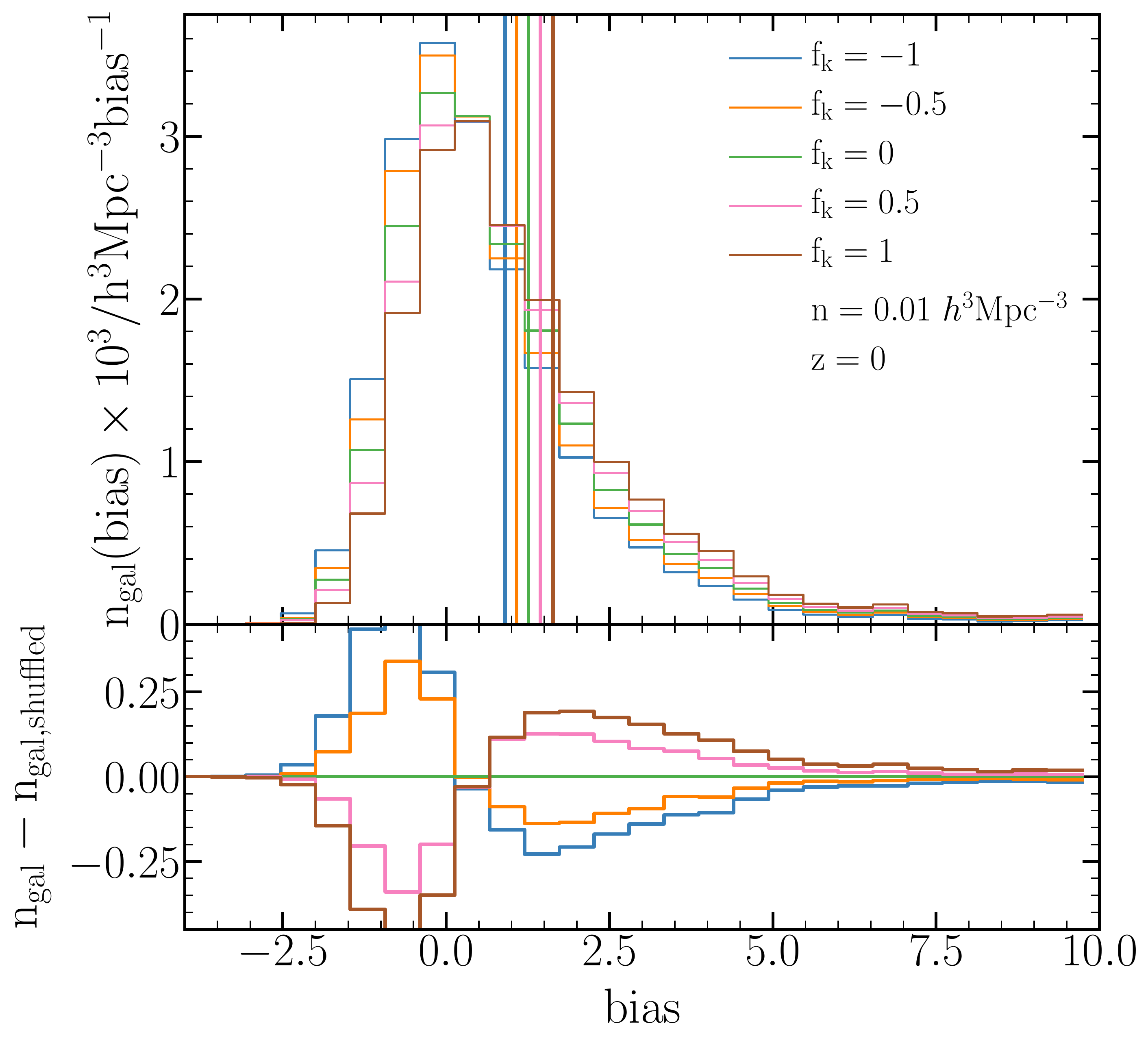}
\caption{Similar to Fig.~\ref{Fig:bias_distro} but for SHAMs with different levels of correlation between the stellar mass and the bias per object, following the procedure explained in \S~\ref{sec:SHAM_AB}. The levels of additional bias are denoted by $f_k$ with $f_k = -1$ perfected anticorrelation between stellar mass and bias, $f_k = -0.5$ moderated anticorrelation, $f_k=0$ no correlation (equivalent to a standard SHAM), $f_k = 0.5$ moderated correlation and $f_k = 1$ complete correlation.}
\label{Fig:fk_hist}
\end{figure}

Since we compare catalogues above a stellar-mass threshold, the previous correlations imply that our samples will display different distribution of biases, clustering amplitudes, and degrees of assembly bias. We can see this in Fig.~\ref{Fig:fk_hist}, which shows for the SHAM samples with varying values of $f_k$, the distribution of biases and the correlation functions relative to their shuffled version. The respective clustering, relative to their shuffled version, is shown in the bottom right panel of  Fig.~\ref{Fig:vpeak_mstell_bias}.

Consistent with our previous discussion, we see that the higher the value of the assembly-bias-correlation parameter, $f_k$, the more preferentially high-bias haloes will be selected, {which results in an increase of a 90\% in the correlation function (compared to the shuffled version).} In contrast, lower $f_k$ preferentially select low-bias haloes, which implies a negative amount of assembly bias reducing the correlation function amplitude by up to 40\%. In turn, $f_k=0$ shows an identical distribution as that of the original catalogues, thus the assembly bias stays at the 15\% level in agreement with the standard SHAM analysis of \cite{ChavesMontero:2016}. 

\subsection{Our model in practice}

\begin{figure}
\includegraphics[width=0.45\textwidth]{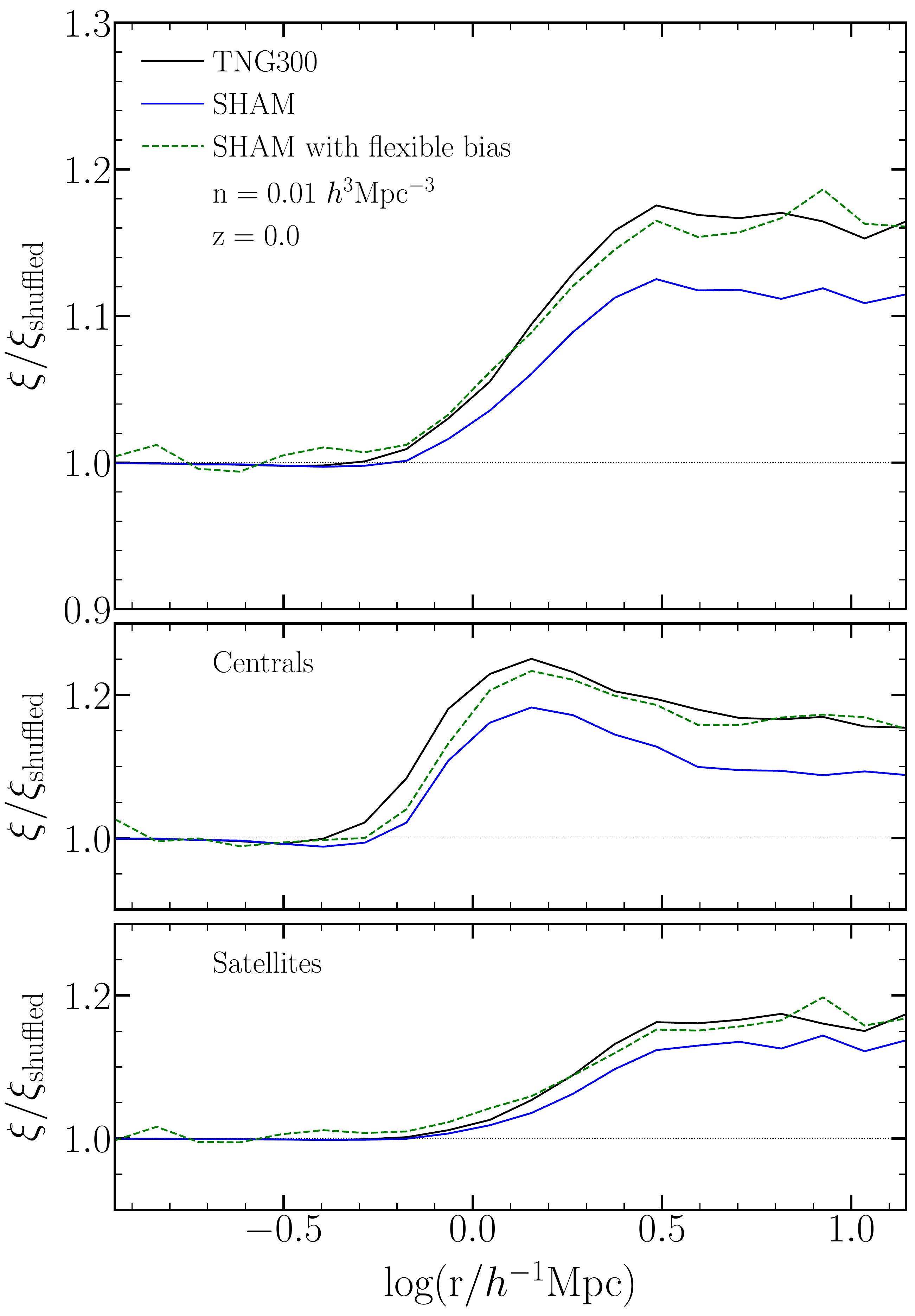}
\caption{(Top) The ratio between the correlation function of the TNG300 and it shuffled run (black solid line) and SHAM and it shuffled run (blue solid line), for a number density of $\rm n=0.01\,\ihMpcC$ selected by stellar mass at $\rm z=0$. The green dashed line shows the prediction of the SHAM with assembly bias, as explained in section \ref{sec:SHAM_AB}. (middle) Same to the top panel, but for the cross-correlation between the central galaxies of the sample and the full galaxy sample. (bottom) Same as the middle panel, but for satellites instead of central galaxies.}
\label{Fig:TNG_SHAMe}
\end{figure}

\begin{figure}
\includegraphics[width=0.45\textwidth]{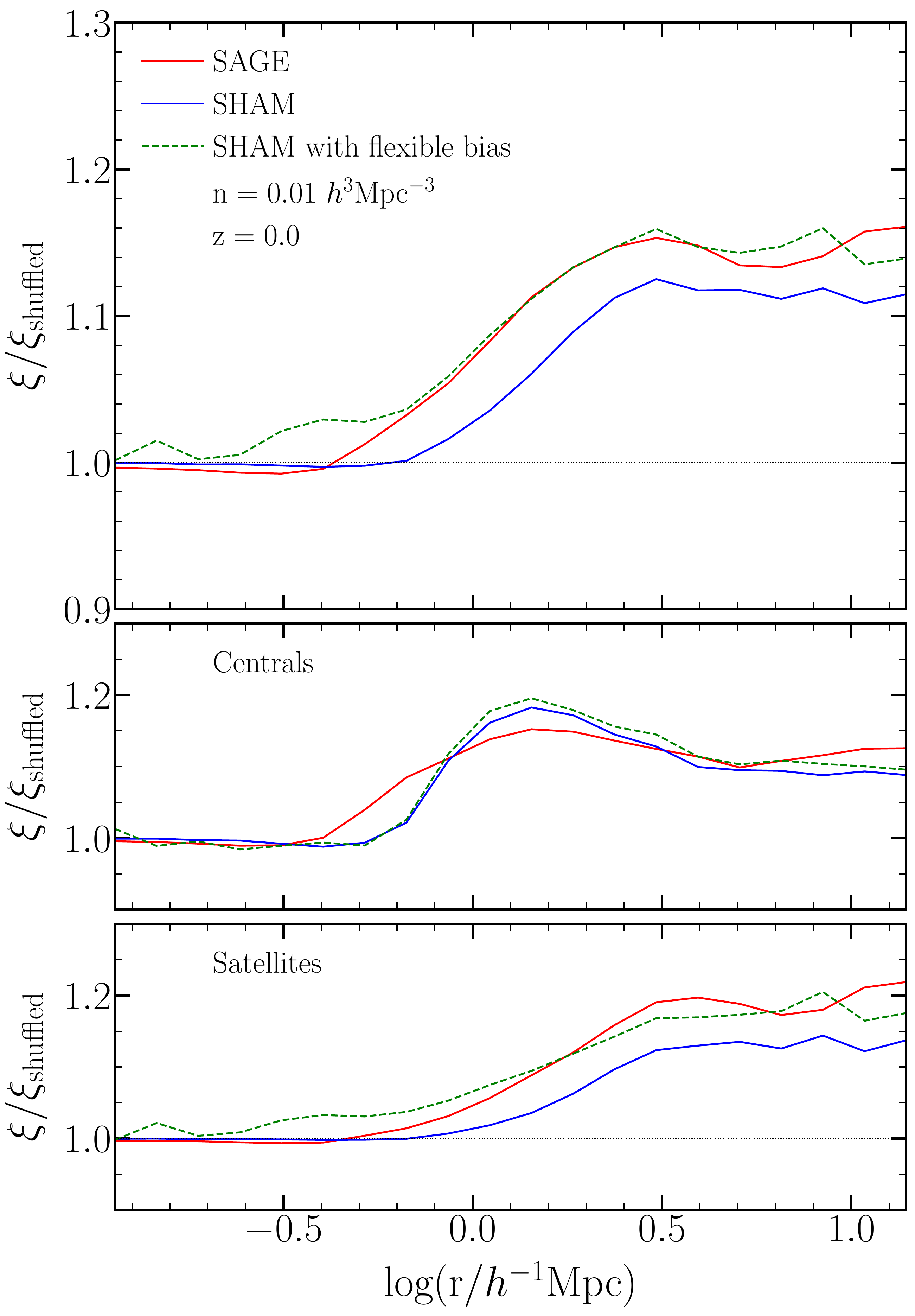}
\caption{Same as Fig~\ref{Fig:TNG_SHAMe}, but for galaxies from SAGE semi-analytical model (solid red lines) instead of the TNG300 hydrodynamic simulation.}
\label{Fig:SAGE_SHAMe}
\end{figure}

We now apply our model and show that with it, SHAM can mimic the magnitude of assembly bias measured in either the TNG300 or SAGE-SAM catalogues.

For this, we first fit the values of $\fkcen$ and $\fksat$ in SHAM that provides the best match to the difference between the original and shuffled distribution in either SAGE or the TNG. Then, we constructed a new catalogue with these values and measure its clustering properties. We chose to use the same SHAM to fit the galaxy assembly bias signal from both, the TNG300 and the SAGE-SAM. This supposes an extra challenge since the SAGE-SAM has a different scatter in its $\vpeak$-stellar mass relation than the TNG300. By using the same SHAM it facilitates the comparison between models, and help us to prove the flexibility of our model.

In Figs.~\ref{Fig:TNG_SHAMe} and Fig.~\ref{Fig:SAGE_SHAMe} we compare the assembly bias signal between the TNG300, SAGE, and our original and new (enhanced) SHAM catalogues. We display the case of a stellar mass-selected sample with a number density of $n=0.01\,\ihMpcC$ at $z=0$. Middle and bottom panels show the assembly bias only for central and satellite galaxies, respectively. The values to mimic the TNG300 clustering are $f_k = $ 0.24 and 0.075 and for SAGE are $f_k = $ 0.01 and 0.19, for centrals and satellites respectively. We notice that $\fksat<\fkcen$ for the TNG300 but  is the other way around for SAGE ($\fksat>\fkcen$). We checked that these relations hold also for the other number densities and redshifts considered in this work.

Overall, we see that our model reproduces very well the amount of assembly bias present in SAM or TNG, both for central and satellite galaxies. It is particularly striking that, although the values of $f_k$ were set to reproduce the large-scale assembly bias, they do also reproduce the scale-dependence of assembly bias on intermediate scales, i.e $1 < r [\hMpc] < 5$. The agreement is particularly remarkable for the case of TNG galaxies. For both central and satellite galaxies, the amplitude and scale-dependence of the galaxy assembly bias coincide to a few percents. The agreement is somewhat poorer for central galaxies in SAGE, especially on intermediate scales. We note that these scales in the correlation function of central galaxies receive an important contribution of ``splashback'' galaxies, thus they are sensitive to the way SAGE  treats them (e.g. \citealt{Zehavi:2019}). In any case, they contribute in a minor way to the full correlation function, whose assembly bias is also reproduced to a few percents in our model. Also, as mentioned on \S~\ref{sec:SHAM}, the scatter of the SHAM is set to mimic the scatter of the TNG300, so we expect a better agreement between those models.  Although not shown in the main body of this paper, we highlight that we find similarly good agreement for all the samples considered here. In Appendix A, we show the respective results.

We remind the reader that having the same assembly bias does not guarantee to have the same correlation function. This is because the SHAM has other limitations besides assembly bias (see \citealt{Smith:2016, Campbell:2018}). In a future work, we plan to use the improvements shown here with other extensions to the SHAM to accurately reproduce the galaxy clustering of more sophisticated galaxy formation models.

\subsection{Galaxy assembly bias in the bispectrum}

Since our model manipulates internal correlations of the catalogue, one might wonder whether other statistical properties of the sample are preserved. To explore this question, we have computed the bispectrum of our original and shuffled catalogues. This quantity is defined as: 

\begin{equation}
B({\bf k_1},{\bf k_2},{\bf k_3}) = \langle \delta({\bf k_1}) \delta({\bf k_2}) \delta({\bf k_3})\rangle \delta_D({\bf k_1} + {\bf k_2} + {\bf k_3}) 
\end{equation}

\noindent where $\delta_D$ is the Dirac's delta. We have considered isosceles (|$\vec{k_1}|=|\vec{k_2}|=|\vec{k_3}|$) and squezed triangular configurations ($|\vec{k_1}|=0.01\ihMpc$, and $|\vec{k_2}|=|\vec{k_3}|$). In particular, the squeezed configuration will test whether the correlations between small scales is responding adequately to fluctuations on larger scales. 
To measure these bispectra we use the publicly available {\sc bskit} code\footnote{https://github.com/sjforeman/bskit} \citep{Foreman:2019}.

\begin{figure}
\includegraphics[width=0.45\textwidth]{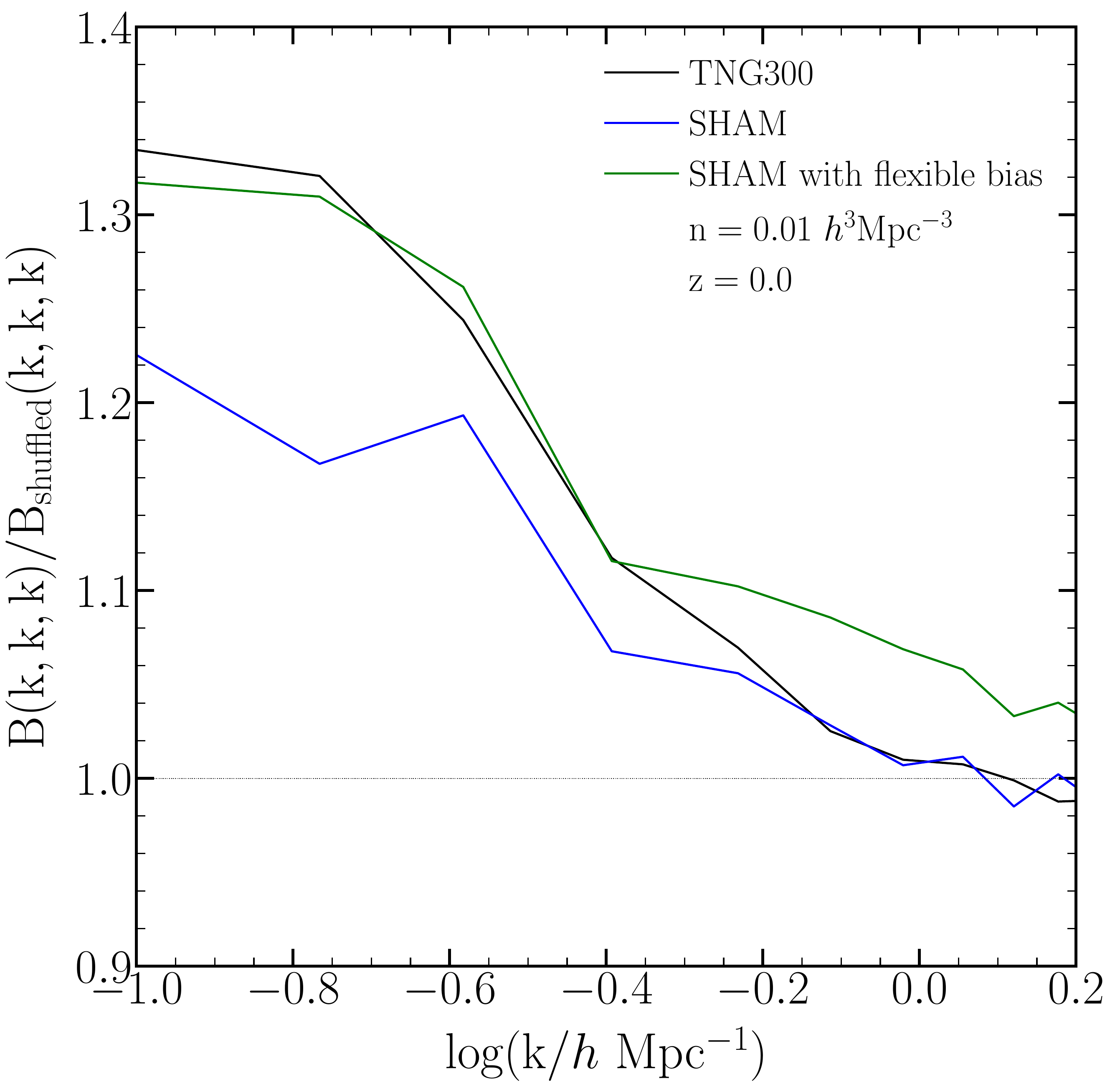}
\includegraphics[width=0.45\textwidth]{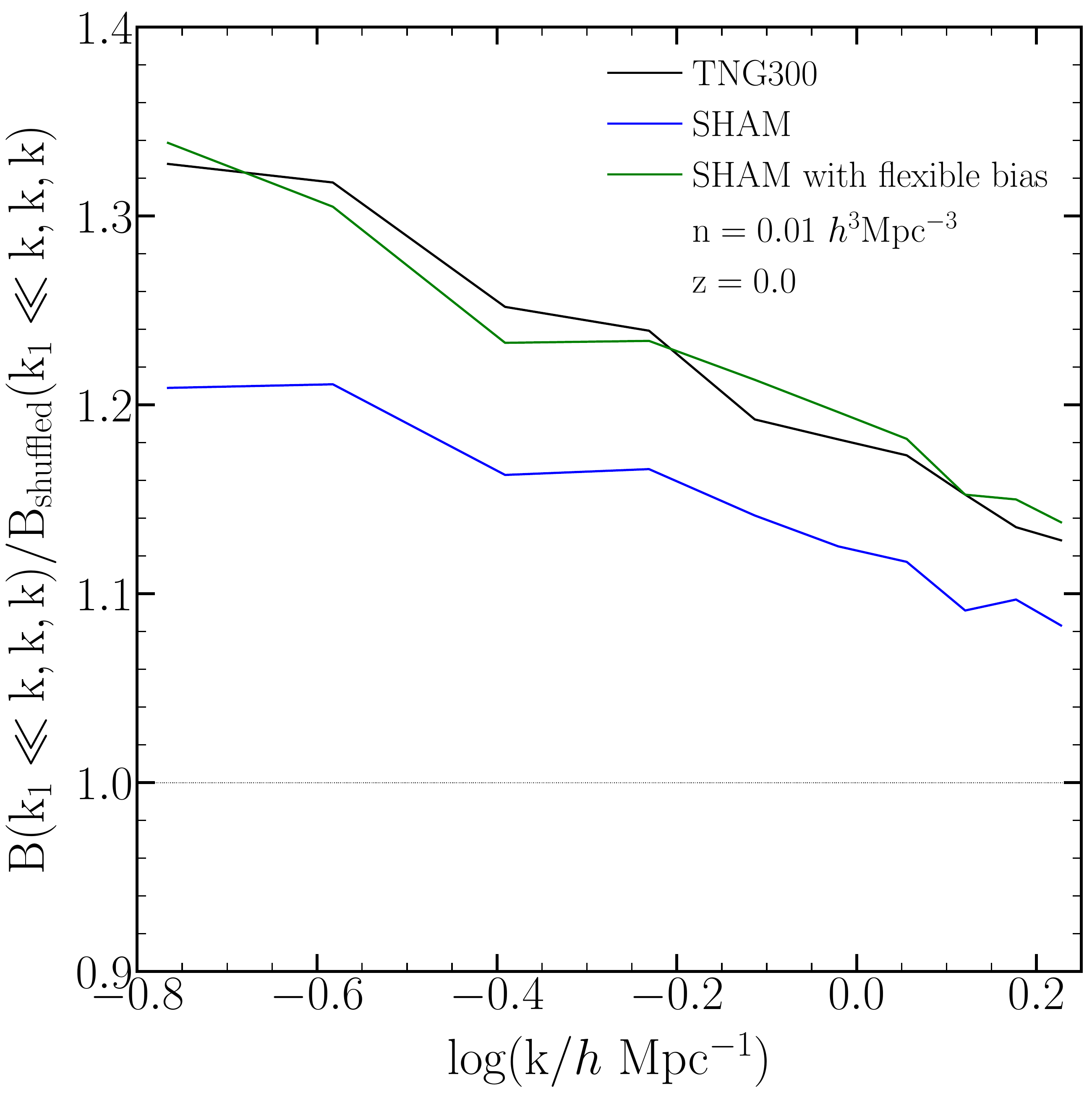}
\caption{
\label{Fig:bispec} The ratio between the galaxy bispectrum to that of its shuffled version. We consider a catalogue of stellar-mass selected galaxies at $z=0$ with a number density of $0.01\ihMpcC$. Top panel shows an isosceles configuration, ($|k_1|=|k_2|=|k3|=k$), whereas the bottom panel shows an squeezed configuration with $k_1 = 0.01\ihMpc$ and $|k_2|=|k_3|=k$. In each panel, black, blue and green lines denote results for the TNG300 simulation, SHAM mocks, and the version of SHAM with a tuneable degree of assembly bias presented in this paper.}
\end{figure}

We show our results in Fig.~\ref{Fig:bispec}, where we display the measured bispectrum in our original catalogues over that in their shuffled counterpart. As in the case of the power spectrum, we can see that SHAM underestimates the amount of galaxy assembly bias in the bispectrum, for both triangular configurations displayed. In contrast, our model with additional assembly bias agrees remarkably well with that measured in the TNG300 galaxies. Although this agreement is primarily a consequence of the agreement in the two-point statistics, we highlight that the values to fit the correlation parameters were set making no reference whatsover to this three-point statistics. In other words, our values for $f_k$ were set to reproduce the effect on the mean large-scale bias of the sample, the bispectrum, however, is also sensitive to higher-order cumulants of the distribution. Thus, there could be infinitely many values of assembly bias in the bispectrum for a given effect in the correlation function. Finally, to our knowledge, this is the first measurement of the effect of galaxy assembly bias in a three-point statistics.

All these cases illustrate the flexibility and accuracy of our technique. While this time we were limited by the volume of the TNG300 simulation, larger simulations would allow for more detailed investigations and possible further improvements.

\section{Summary and Conclusions}
\label{sec:Conclusions}

In this paper, we studied the behaviour of galaxy assembly bias of various samples with different number densities, redshifts, and selection criteria. We use the classic definition of galaxy assembly bias introduced by \cite{Croton:2007} that is the difference in the large scale clustering of galaxies at a fixed halo mass due to correlations with the assembly history and other properties of host haloes and can be measure by comparing the galaxy clustering of a sample with the one of its shuffled counterpart. We consider three different galaxy models. Specifically, the TNG300 simulation, a state-of-the-art magneto-hydrodynamic simulation of 205 $\hMpc$; SAGE, a  state-of-the-art semi-analytical model of galaxy formation; and subhalo abundance matching built with $\vpeak$. These three models were performed over the same simulated cosmic volume, which enables a precise comparison.

Below we summarise the main results of this work:

\begin{itemize}
  \item Quantifying the redshift evolution and dependence with number density and for galaxies selected by stellar mass and SFR, we find that all the models feature different amplitude for the galaxy assembly bias. The differences were particularly evident for star-forming samples. (Figs.~\ref{Fig:shuffle} \&\ ~\ref{Fig:shuffle_SFR}).

  \item We found that the evolution with redshift and number density of the galaxy assembly bias are similar for SAGE and SHAM. This is in agreement with previous results from the literature (eg. \citealt{C19}), but they are different to that of the TNG300. Based on this we argued that, while the presence of galaxy assembly bias is part of the current galaxy formation theory, its amplitude and behaviour is strongly model-dependent. (fig.~\ref{Fig:bias_ev}).

  \item By looking at the individual large-scale bias of the galaxies, we showed that galaxy assembly bias is equivalent to how different selection criteria and physics modelled preferentially select locations with different large-scale bias. While not surprising, this perspective can improve the way we focus our efforts on understanding the origin of galaxy assembly bias and in the creation of mocks with galaxy assembly bias. (fig.~\ref{Fig:bias_distro}).

  \item We developed a method to model assembly bias in SHAM. The method works by re-ordering the galaxies keeping constant their $\vpeak-M_{{\rm stell}}$ relation and its distribution of satellites. We find that by maximising or minimising the correlation with the bias, we are able to modify the large scale clustering by a factor of 3 (meaning $\sim 70\%$ in the differences of the bias of the sample). (fig.~\ref{Fig:vpeak_mstell_bias}).

  \item We used our SHAMs extended with assembly bias to reproduce the level of galaxy assembly bias in the TNG300 or SAGE catalogues. These mocks can be used to create catalogues with a fixed assembly bias signal (e.g. in case we want the same level of assembly bias of a hydrodynamic simulation or a SAM) or we can let it free when interpreting the observed galaxy clustering (Figs.~\ref{Fig:TNG_SHAMe} \&\ ~\ref{Fig:SAGE_SHAMe})

This new SHAM model presents an upgrade of previous improvements to the standard model by mimicking the assembly bias signal in a flexible way, based on an environmental halo property (not concentration) and, thanks to the imposition of keeping the satellite fraction constant, that only add galaxy assembly bias while keeping the clustering amplitude fixed. We anticipate our extended SHAM catalogues to have various applications. First, they could help in designing new observational tests to measure galaxy assembly bias in galaxy surveys. This is done by creating mocks with different amplitudes of assembly bias and fitting the total galaxy clustering from observations (similar to the approach of \citealt{Salcedo:2020}). This will constrain the level of assembly bias necessary to properly reproduce the correlation function, indicating us the amount of assembly bias from the observed universe. With the same idea, it can be used to put constraints on the maximum level of assembly bias possible and studying its origin, by looking at how the galaxy assembly bias signal evolves with redshift and number density. Furthermore, it can help to explore the degeneracy between cosmological parameters and galaxy formation physics (e.g. following the procedure shown in \citealt{C20a}). This could be particularly relevant in current searches for signature of modified gravity, where there exists additional correlations between galaxy properties and large-scale densities. The model presented here is simple enough so it can be implemented without the need of extensive and complex calculation, not only with the ``object-by-object'' bias but any secondary halo property. In a recent work \citep{C20c}, we develop a new extension of the traditional SHAM model that includes a novel treatment of orphan galaxies, tidal disruption for the satellite galaxies, star formation rate predictions and a flexible level of galaxy assembly bias using the model presented in this paper.

\appendix

\section{Additional results}

In this appendix, we provide a comparison between the assembly bias in our catalogues to that obtained from our SHAM mocks extended with assembly bias. Fig.~\ref{Fig:TNG_SHAMe_All} shows the  ratio between the correlation function of the TNG300 and it shuffled run and SHAM and it shuffled run (similar to fig.~\ref{Fig:TNG_SHAMe}) for $n=0.01,\ \&\ 0.00316\,\ihMpcC$ at $z=0$, $z=0.5$ and $z=1$. In general, we find a good agreement between our extended SHAM and the level of galaxy assembly bias predicted by the TNG. The same predictions but for SAGE are shown in Fig.~\ref{Fig:SAGE_SHAMe_All}. While the agreement between our SHAM with assembly bias and the galaxy formation model is not as good as with the TNG300, it still reproduces quite well the galaxy assembly bias signal.

\begin{figure*}
\includegraphics[width=0.3\textwidth]{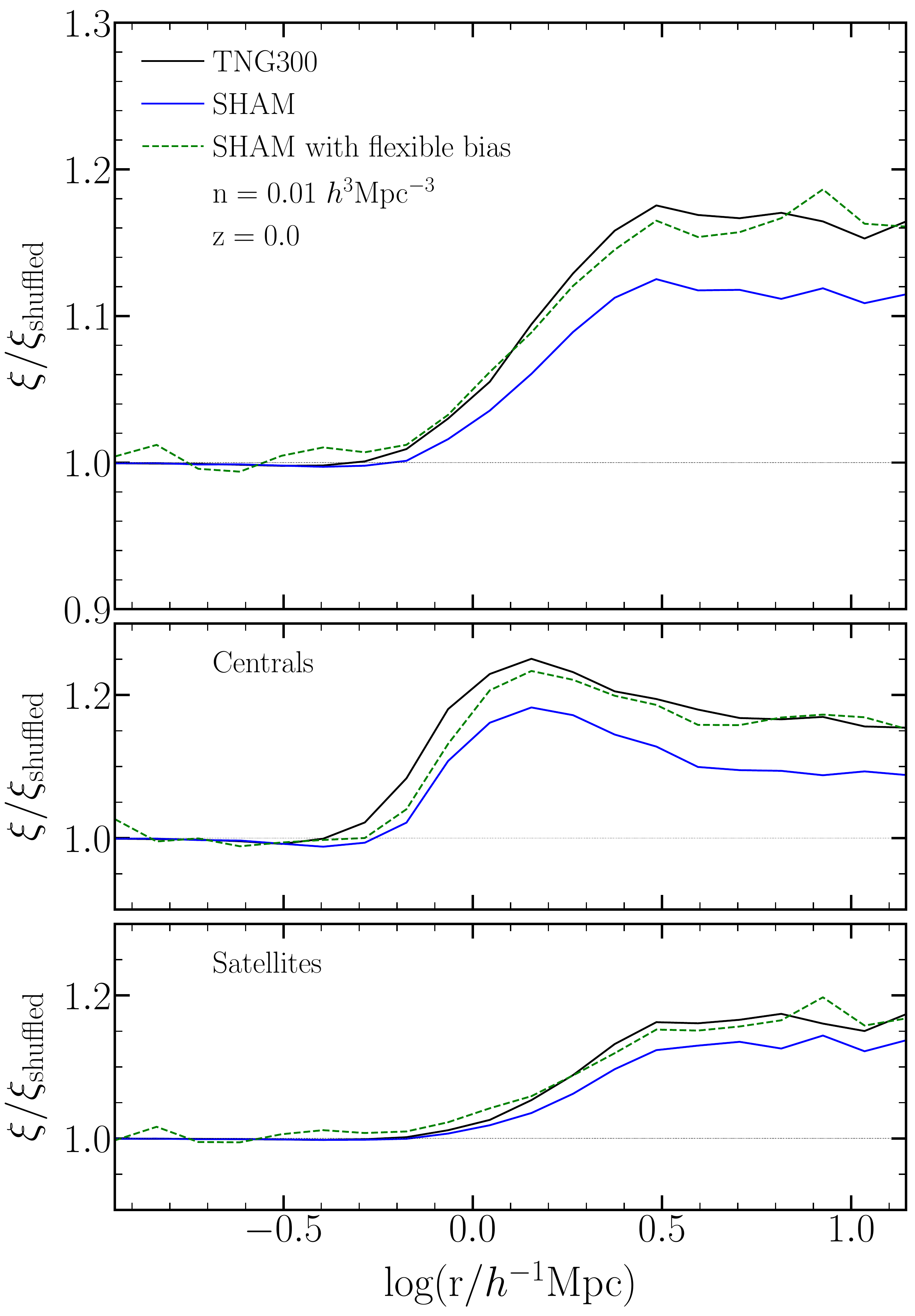}
\includegraphics[width=0.3\textwidth]{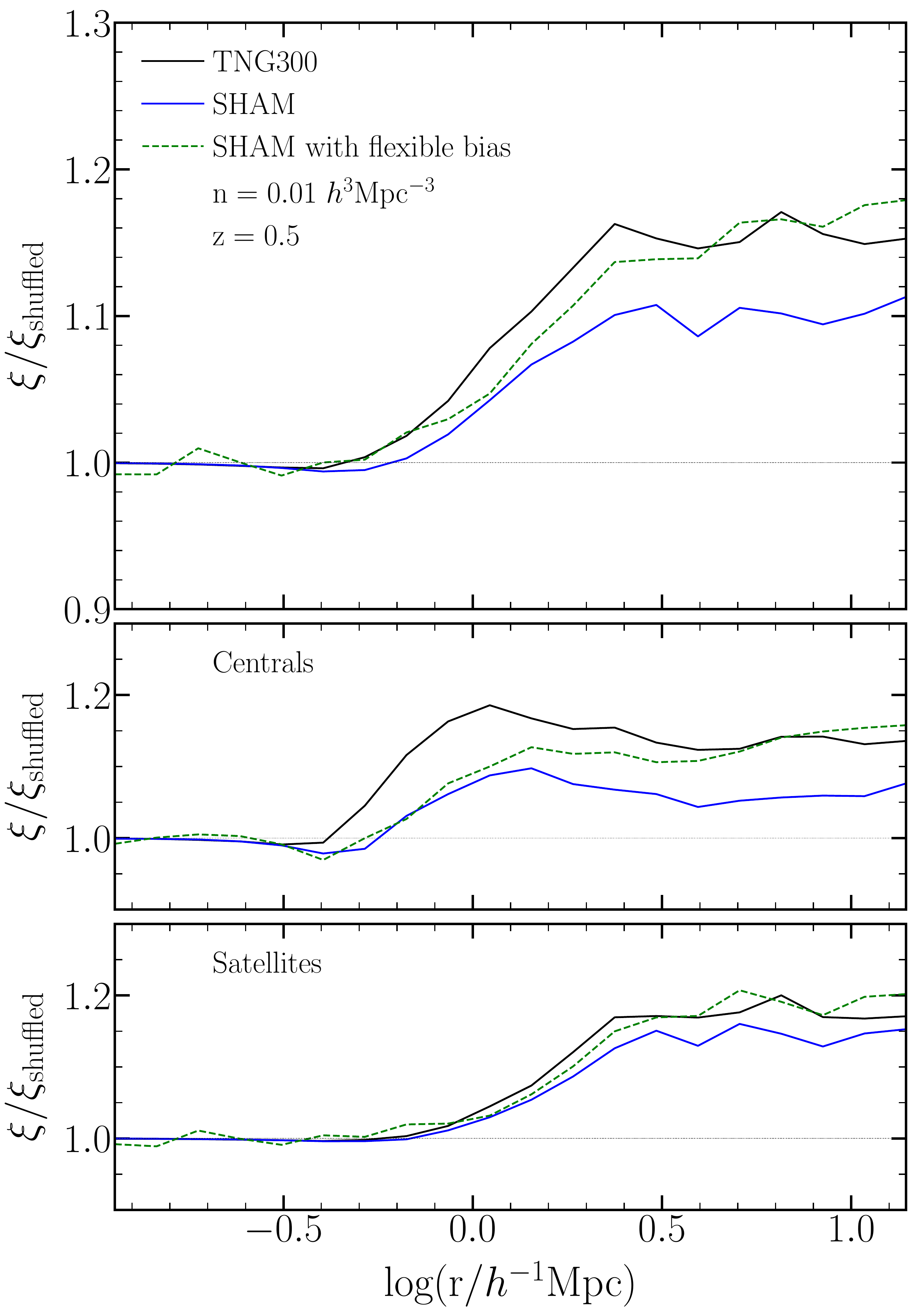}
\includegraphics[width=0.3\textwidth]{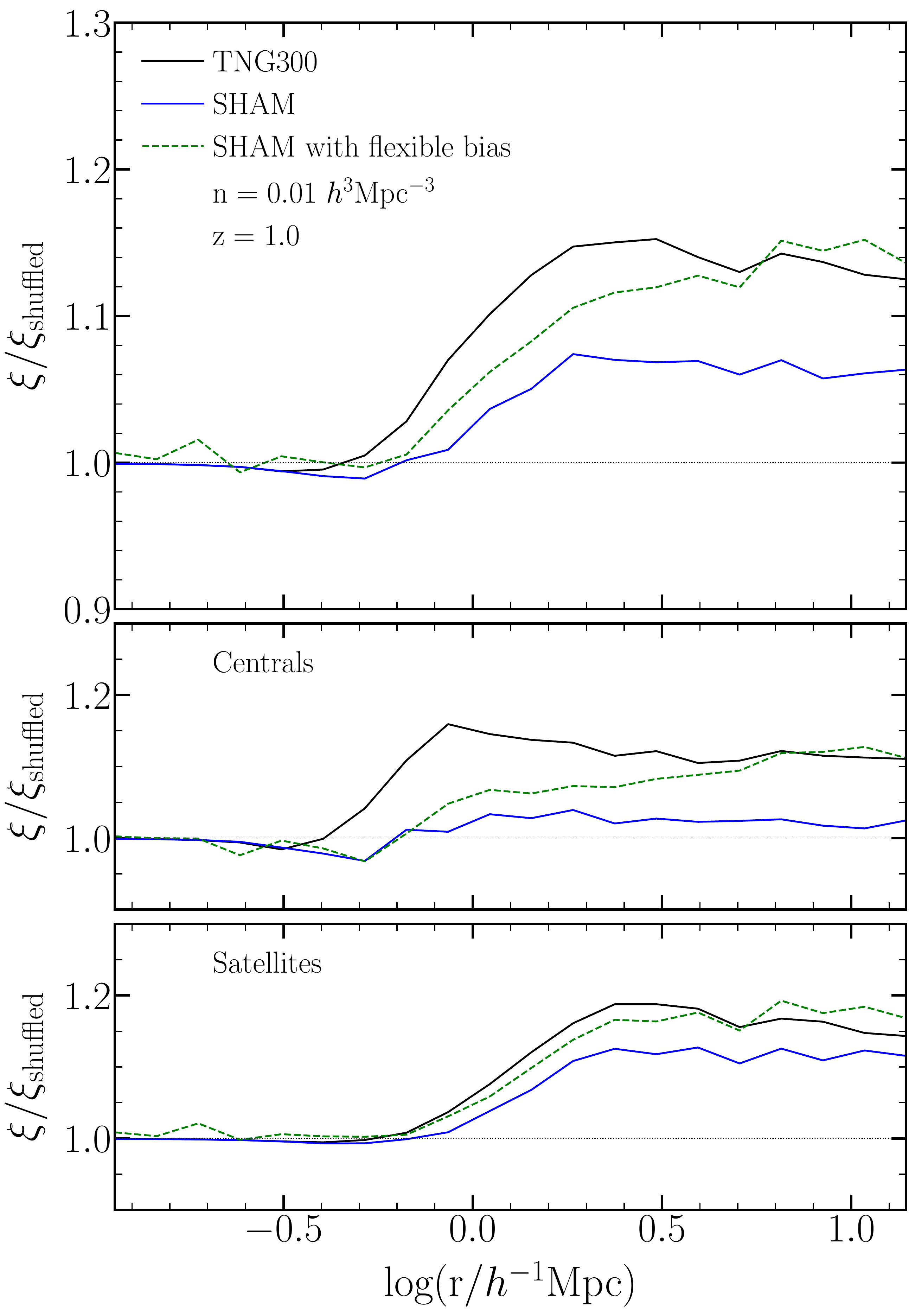}
\includegraphics[width=0.3\textwidth]{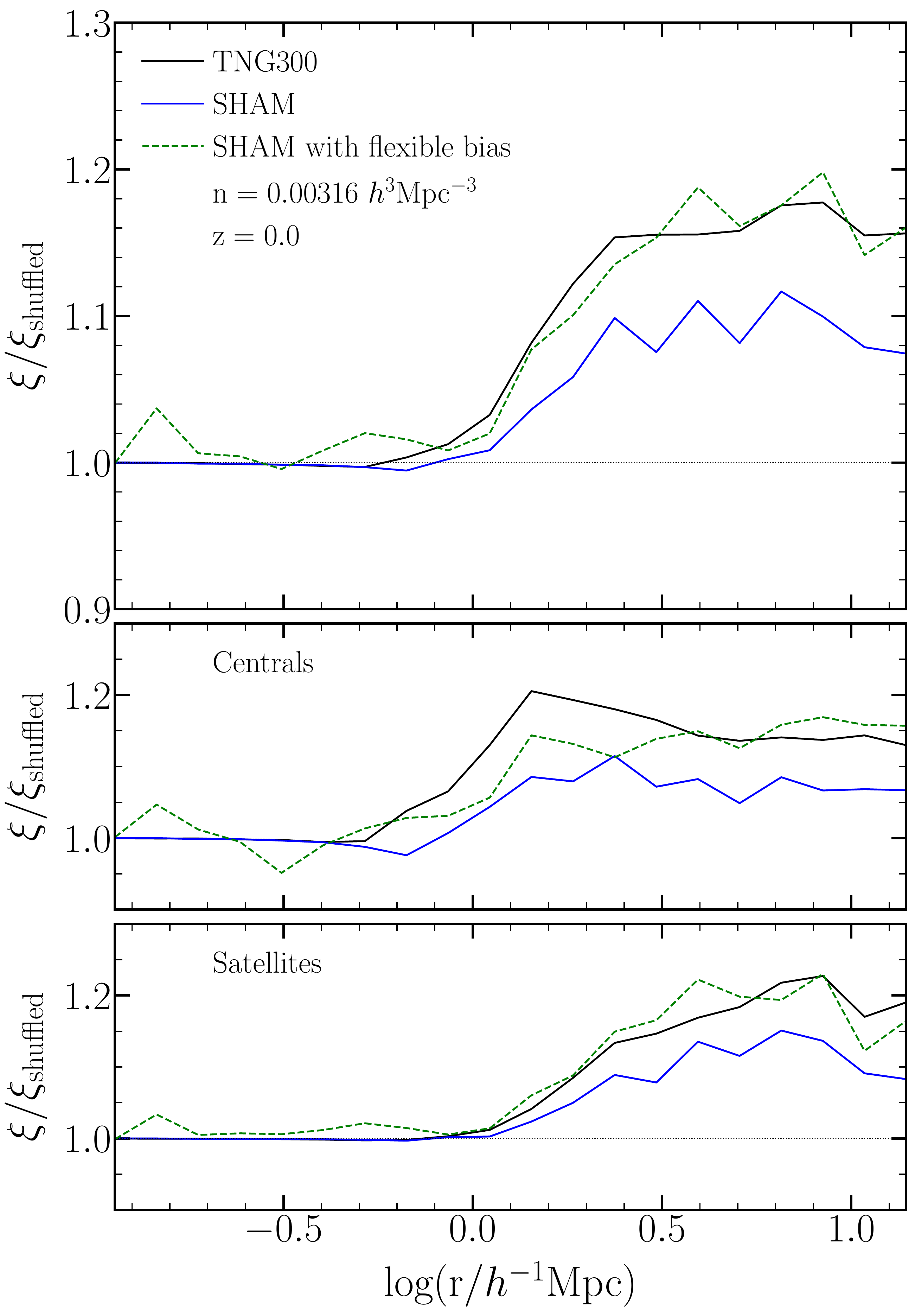}
\includegraphics[width=0.3\textwidth]{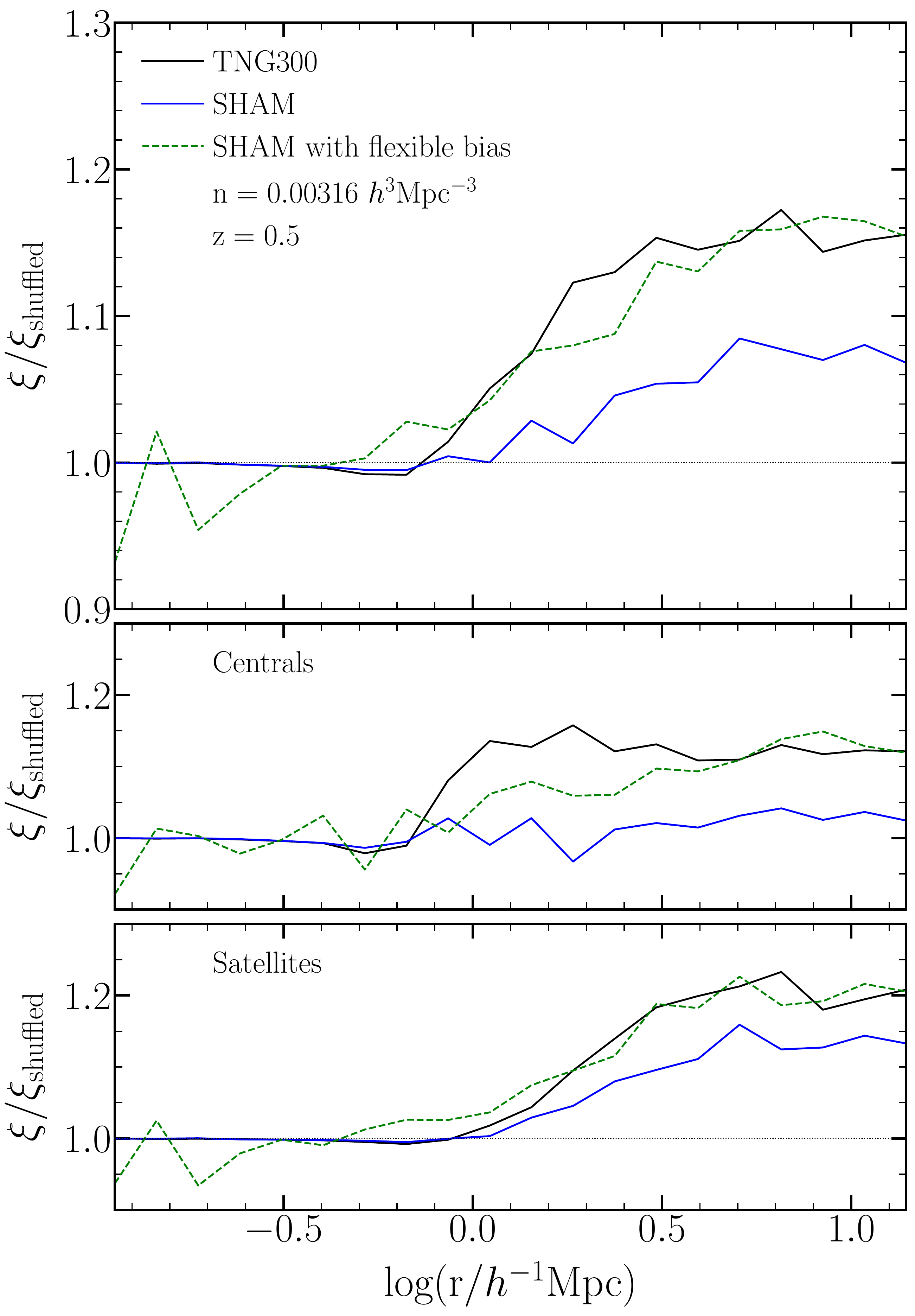}
\includegraphics[width=0.3\textwidth]{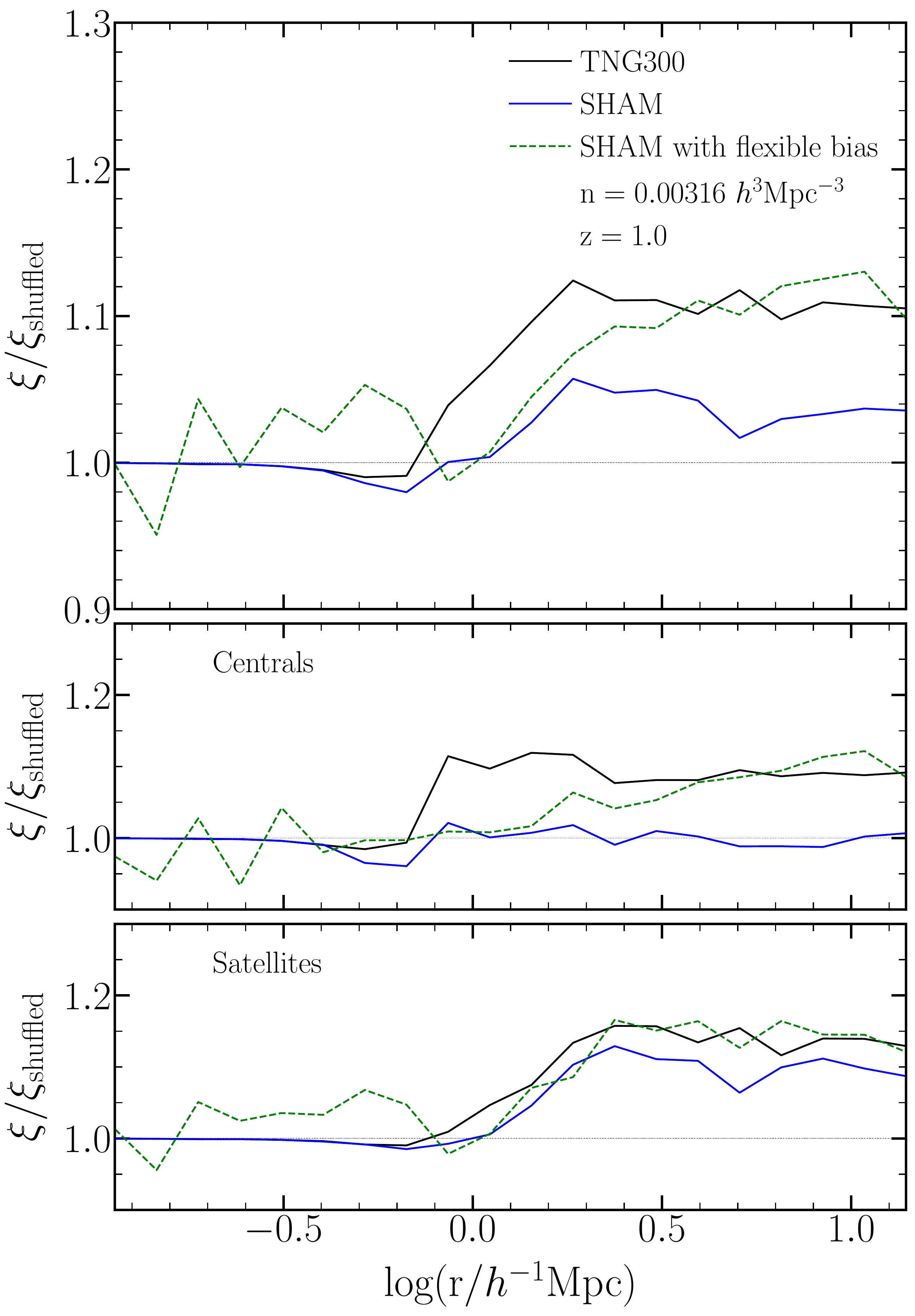}
\caption{Same as Fig.~\ref{Fig:TNG_SHAMe}, but for $n=0.01,\ \&\ 0.00316\,\ihMpcC$ at $z=0$, $z=0.5$ and $z=1$ as labeled.}
\label{Fig:TNG_SHAMe_All}
\end{figure*}

\begin{figure*}
\includegraphics[width=0.3\textwidth]{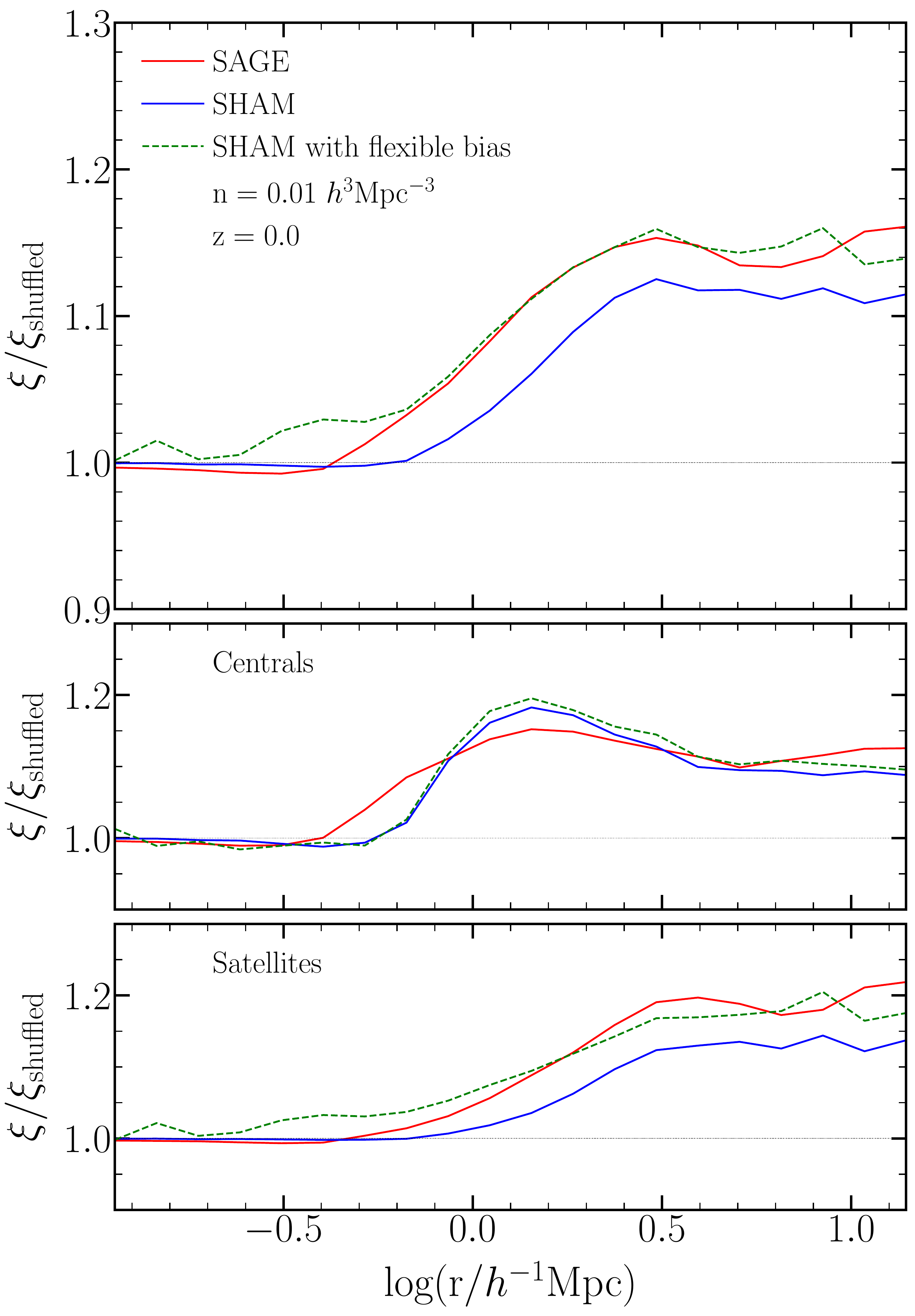}
\includegraphics[width=0.3\textwidth]{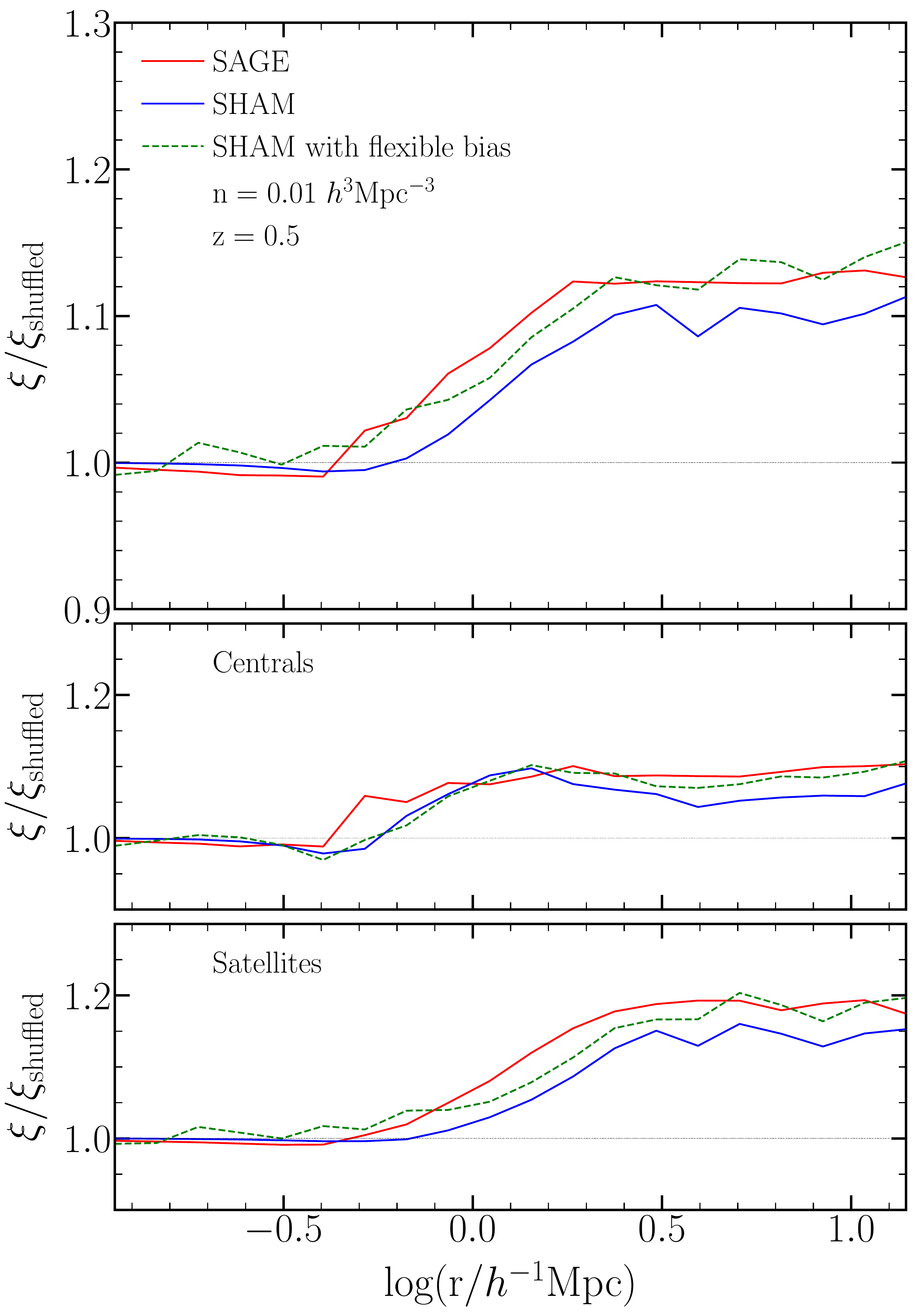}
\includegraphics[width=0.3\textwidth]{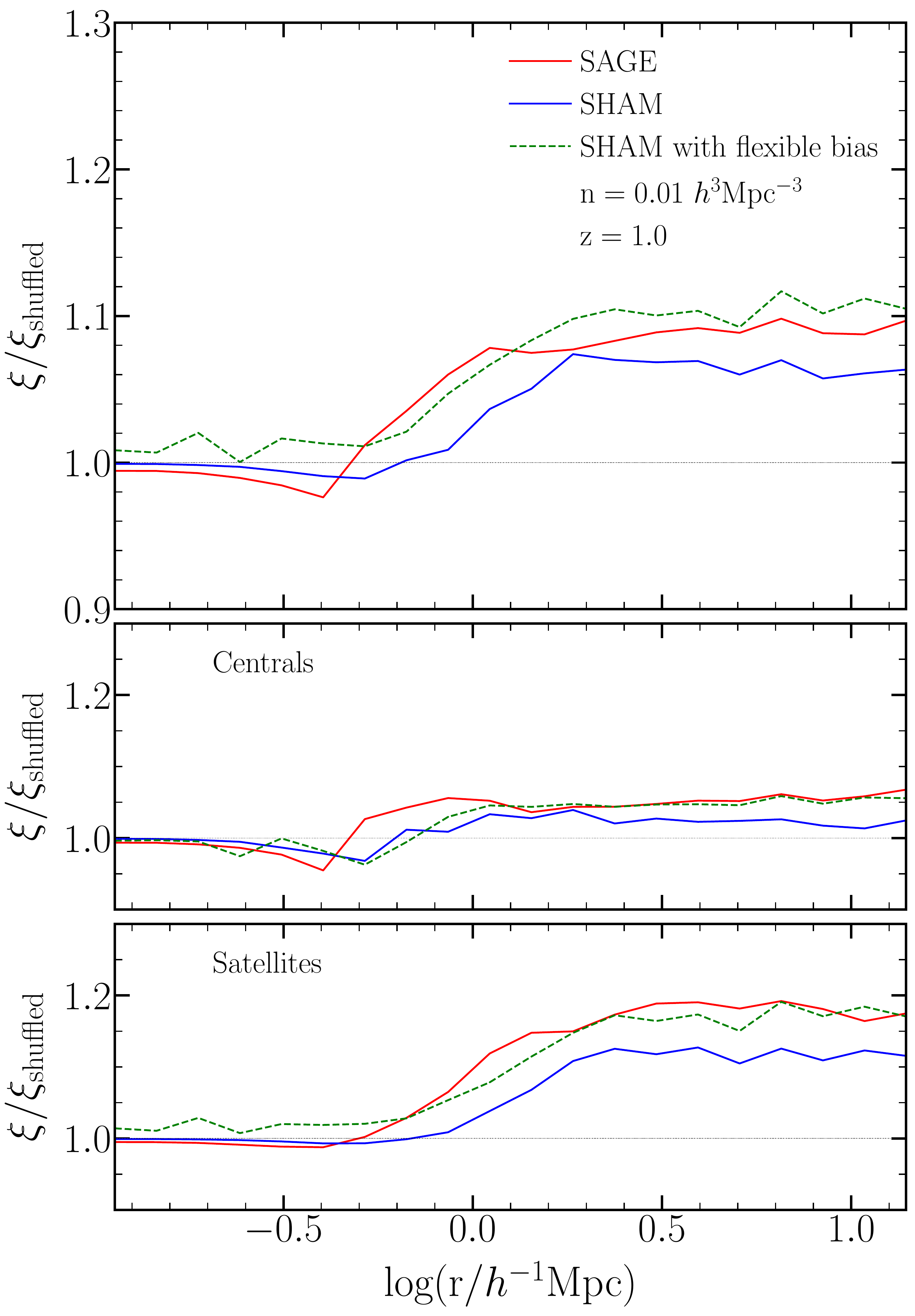}
\includegraphics[width=0.3\textwidth]{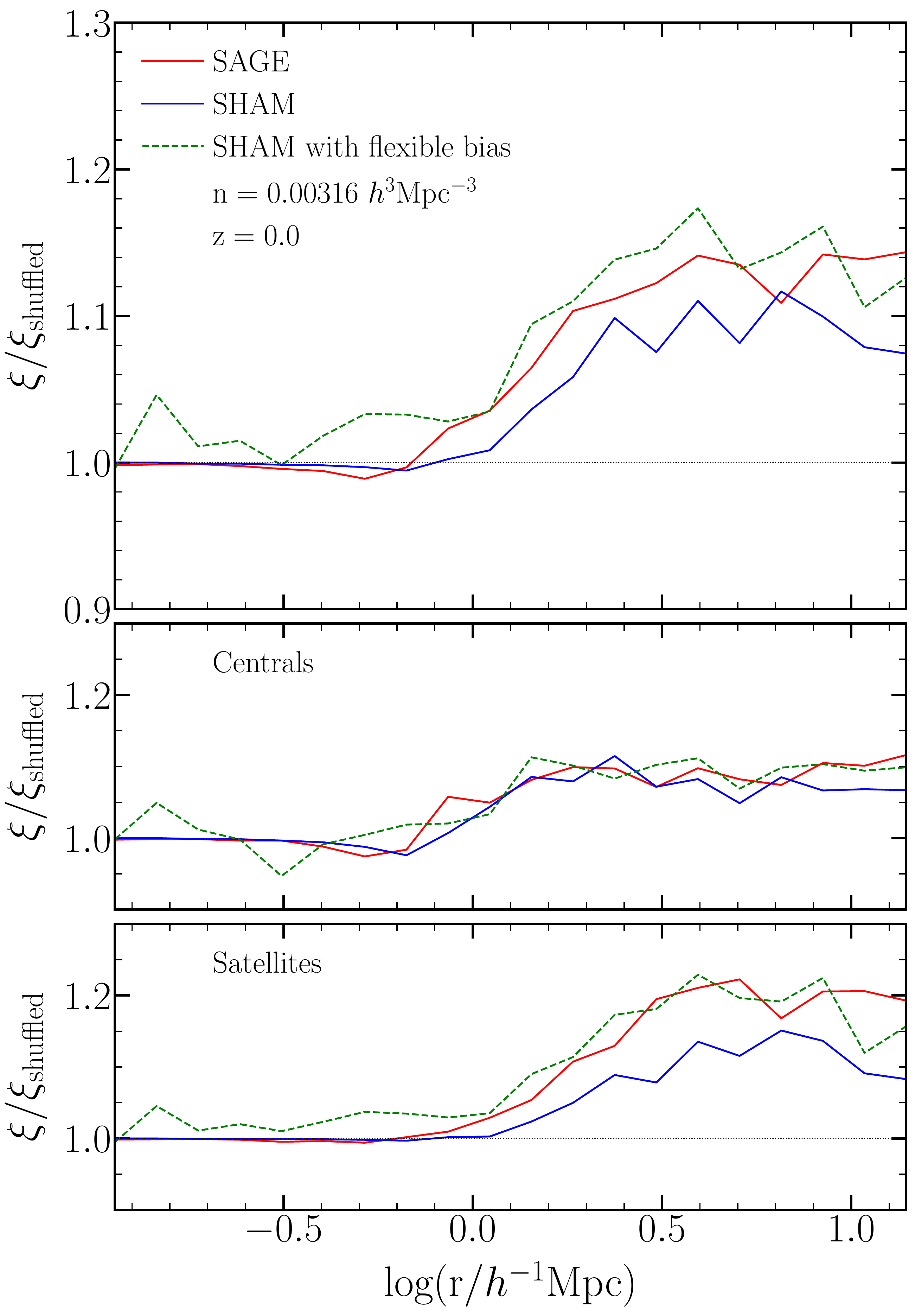}
\includegraphics[width=0.3\textwidth]{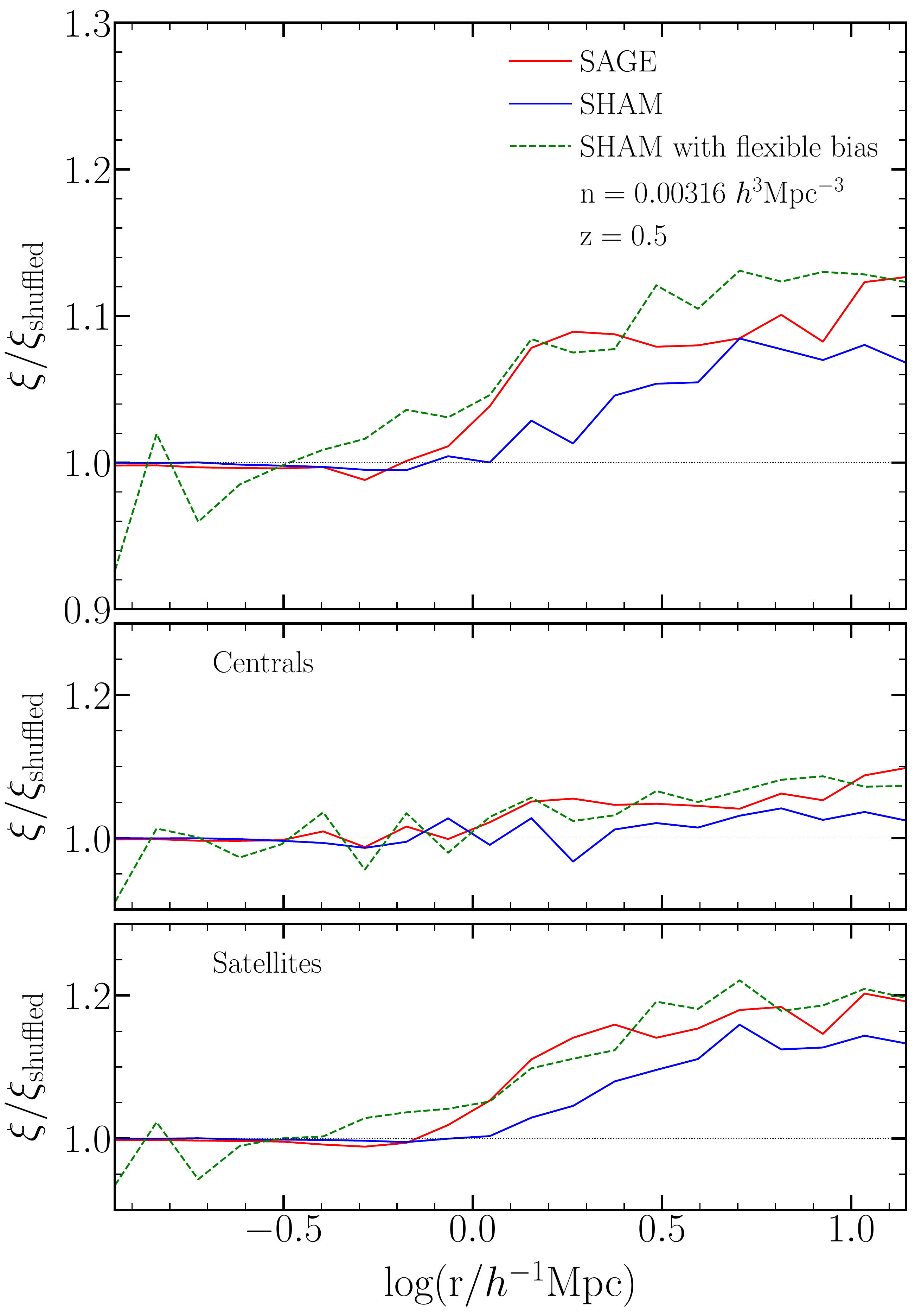}
\includegraphics[width=0.3\textwidth]{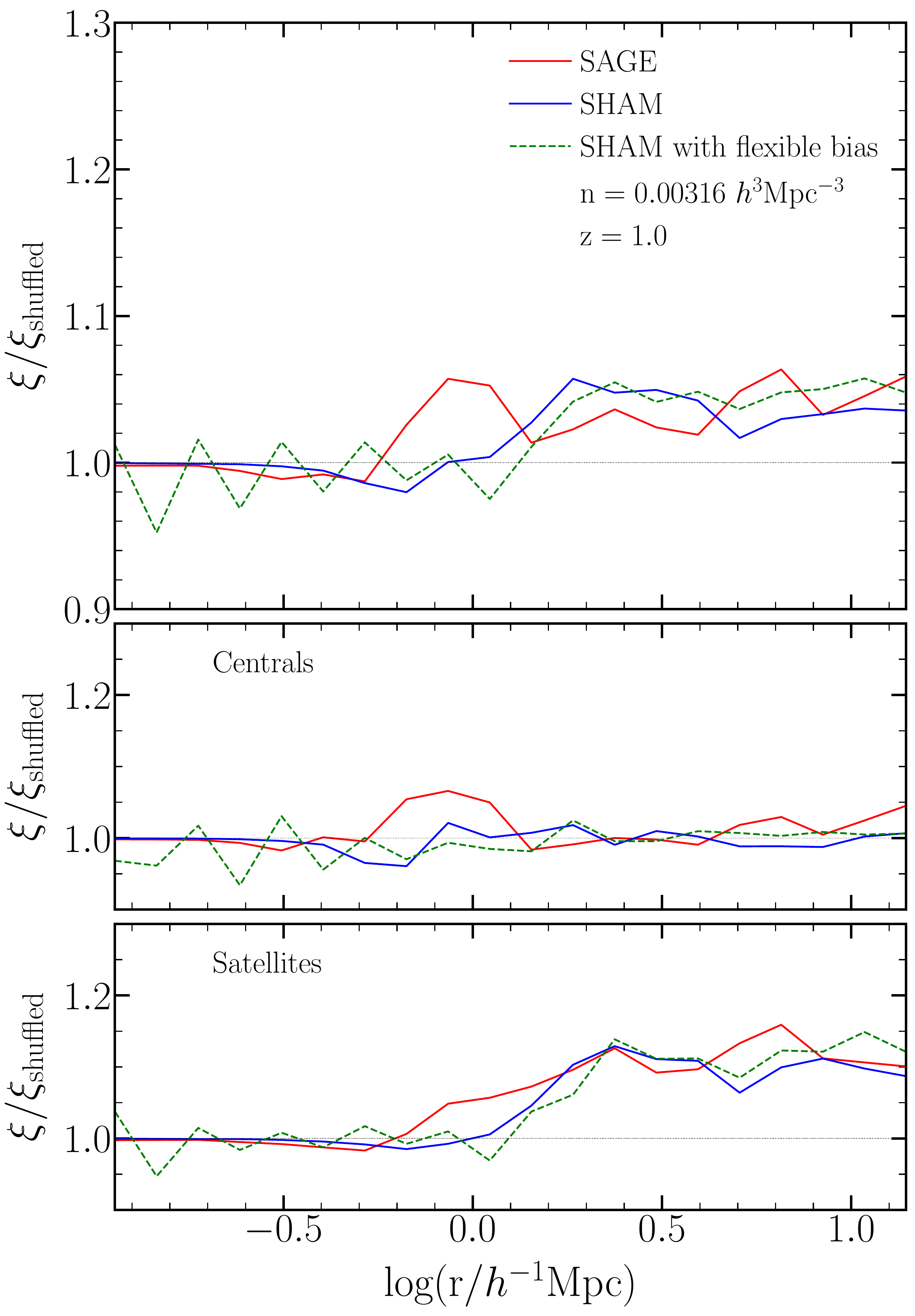}
\caption{Same as Fig.~\ref{Fig:SAGE_SHAMe}, but for $n=0.01,\ \&\ 0.00316\,\ihMpcC$ at $z=0$, $z=0.5$ and $z=1$ as labeled.}
\label{Fig:SAGE_SHAMe_All}
\end{figure*}

\end{itemize}

\section*{Data availability}
The data underlying this article will be shared on reasonable request to the corresponding author.

\section*{Acknowledgements}

We would like to thank the anonymous reviewers for their useful suggestions and comments.
We thanks useful comments from Aseem Paranjape, Antonio Montero-Dorta, Idit Zehavi, Maria Celeste Artale, Nelson Padilla \& Simon White.
The authors acknowledge the support of the ERC Starting Grant number 716151 (BACCO). SC acknowledges the support of the ``Juan de la Cierva Formaci\'on'' fellowship (FJCI-2017-33816). The authors thankfully acknowledge the computer resources at MareNostrum and the technical support provided by Barcelona Supercomputing Center (RES-AECT-2019-2-0012)".

\bibliography{Biblio}

\bsp	
\label{lastpage}
\end{document}